\documentclass[10pt, prd,twocolumn, showpacs,superscriptaddress,notitlepage]{revtex4-1}

\usepackage[a4paper, margin=0.6in]{geometry}
\usepackage{setspace}

\usepackage{amssymb}
\usepackage{amsmath}

\usepackage{graphicx}

\usepackage{color}
\definecolor{darkblue}{rgb}{0,0,1}
\definecolor{darkgreen}{rgb}{0,.4,0}
\definecolor{deepred}{rgb}{.5,0,0}
\usepackage[colorlinks,linkcolor=darkblue,urlcolor=deepred,citecolor=darkgreen,unicode]{hyperref}

\begin{document}
\title{Transmission and reflection of phonons and rotons\\ at the superfluid helium-solid interface.}

\author{I. N. Adamenko}
\email{i.n.adamenko@mail.ru}
\author{K.E. Nemchenko}
\affiliation{Karazin Kharkov National
University, 4 Svobody Sq., Kharkov 61077, Ukraine}
\author{I. V. Tanatarov}
\email{igor.tanatarov@gmail.com}
\affiliation{Akhiezer Institute for Theoretical Physics,
        NSC KIPT of NASU, Academicheskaya St. 1, Kharkov, 61108, Ukraine.}

\begin{abstract}
We solve the problem of  the transmission  and reflection of  phonons
and rotons
at the interface between superfluid helium and a solid,  for all
angles of incidence
  and in both directions. A consistent solution of
the problem is presented which allows us to rigorously describe  the
simultaneous
creation of phonons, $R^-$, and $R^+$ rotons in helium by either a
phonon from the solid or a helium quasiparticle incident on the
interface. The interaction of
all $HeI\!I$ quasiparticles with the interface, and their transmission,
reflection and conversion into each other, is described in a unified way.
The  angles of propagation and the probabilities of creating
quasiparticles are  obtained  for all cases. Andreev reflection of helium
phonons and rotons is  predicted. The energy flows through the
interface due to phonons, $R^-$, and
$R^+$ rotons are  derived.  The  small contribution of the $R^-$ rotons
is due to the small  probability  of  an $R^-$ roton being created by a
phonon in the solid, and vice  versa. This explains the failure to
directly create beams of $R^-$
rotons prior  to  the  experiments  of Tucker and  Wyatt in 1999. New
experiments
for creating $R^-$ rotons, by beams of high-energy phonons (h-phonons), are
suggested.
\end{abstract}


\maketitle

\section{Introduction}

\begin{figure}[bht]
\begin{center}
\includegraphics[viewport=93 0 505 310, width=0.45\textwidth]{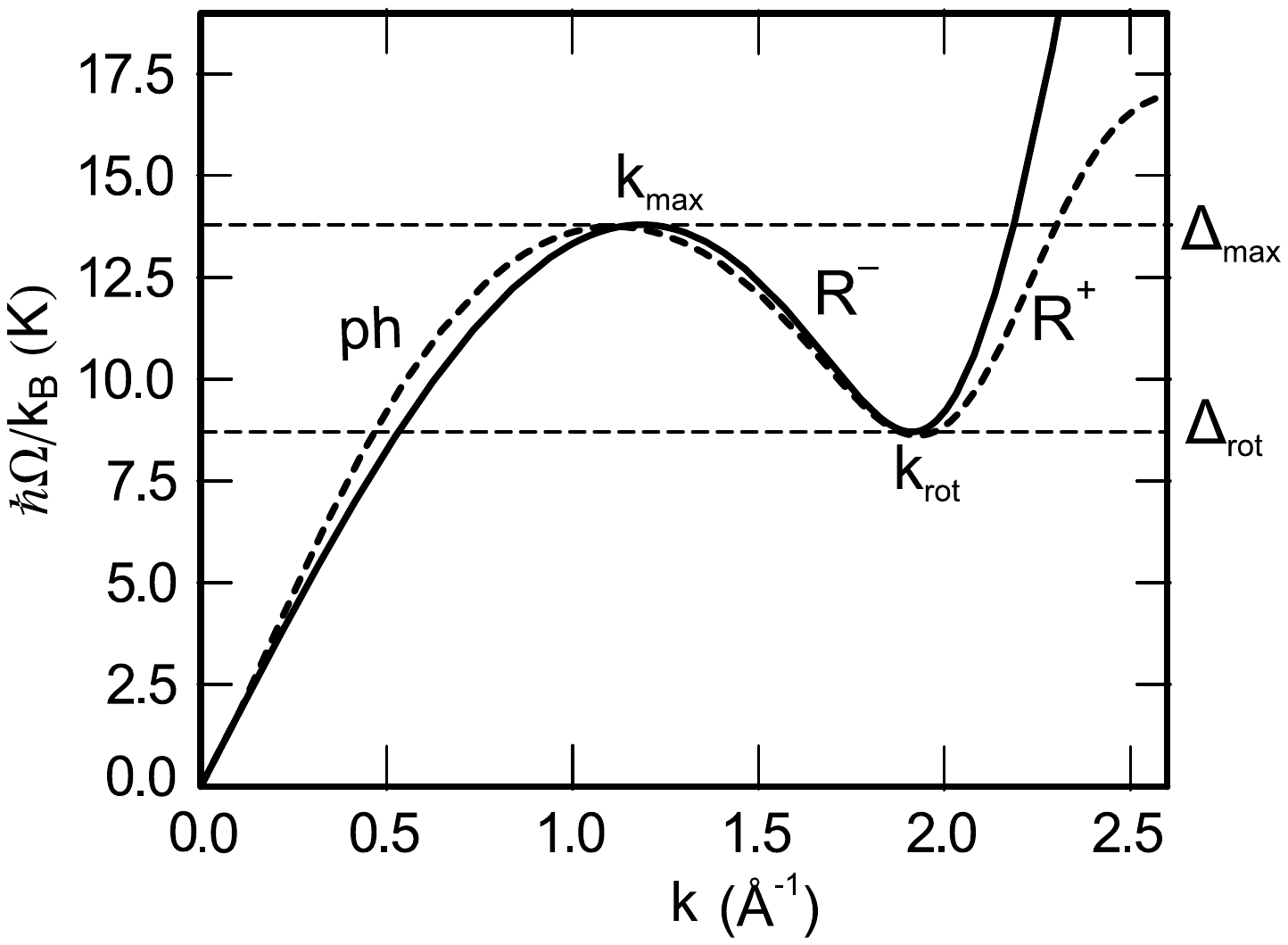}
\end{center}
\parbox{0.45\textwidth}
	{\caption{\label{Fig1} The solid line is $\Omega(k)$ from (\ref{OmegaRot}) for $s\!=\!230.7\,m/s$,  $k_g\!=\!1.9828 {\AA}^{-1}$, and $\lambda\!=\!-0.9667$; the dashed line is the measured dispersion curve of superfluid helium \cite{neutron} at the saturated vapour pressure.}}
\end{figure}

Many physical properties of continuous media at low temperatures can be
described in terms of quasiparticles. The quasiparticles of superfluid helium
  are called phonons, $R^-$ rotons,
and $R^+$ rotons. They have a non-monotonic dispersion curve and the
$R^-$ rotons
  have a negative
group velocity, i.e. their momentum is directed opposite to the group
velocity, see Fig.\ref{Fig1}.
The phonons and rotons are observed in many experiments, such as in neutron
scattering in helium \cite{neutron} and in the direct experiments
\cite{exp1,exp2}
where beams of superfluid helium quasiparticles are created by a
heated solid. The
quasiparticles propagate in the helium, and interact and reflect
from different surfaces. Also  they quantum evaporate helium atoms
from the free surface.
These have been investigated both experimentally and theoretically
(see for example
\cite{phononpulses1,phononpulses2,phononpulses3,phononpulses3a,phononpulses4}).
Interestingly,
$R^-$ rotons were not detected in direct experiments until 1999, when they
were finally created by a specially
constructed source \cite{exp2}. They were observed by quantum evaporation.
  All the earlier attempts to create $R^-$, with ordinary solid
heaters,  were unsuccessful.

The problem of the interaction of  rotons in superfluid helium with
interfaces,
their reflection, transmission, and mode change,  was first considered in
\cite{Bowley}. However, the method used there did not take into account the
simultaneous creation of phonons, $R^+$ and $R^-$ rotons  by a phonon
in the solid
incident on the interface, and it could not distinguish between the $R^+$
and $R^-$ rotons. Later in \cite{PRB} it was shown that $R^-$ rotons cannot be
created at the interface with a solid by a phonon from the solid,
provided we can
neglect the possibility of creation  of the other
quasiparticles in the same process at the interface. In the current work a
consistent solution is introduced, which allows us to
rigorously solve the problem of
the simultaneous creation of phonons and rotons. Also it describes in a
unified way, the interaction of all $HeI\!I$ quasiparticles with the interface:
their transmission, reflection and conversion into each other. These are the
fundamental elementary processes that determine the heat exchange
between $HeI\!I$
and a solid, and the associated  phenomena, such as the Kapitza
temperature jump
(see for example \cite{Khalatnikov}). We investigate all these phenomena. The
probability of creation of each quasiparticle at the interface is derived for
all cases. The failures of attempts to detect $R^-$ rotons prior to experiments
\cite{exp2} is explained, and predictions are made for new experiments on the
interaction of phonons and rotons with a solid and the creation of $R^-$ rotons
at the interface by high energy phonons (h-phonons).

We describe superfluid helium with its distinctive dispersion relation
$\Omega(k)$, with the maxon maximum and roton minimum, within the framework of
the theory developed in \cite{PRB}. The  quantum  fluid  is considered as a continuous medium at all
length  scales.  This  model  is based on the fact that the thermal de
Broglie  wavelength  of  a  particle  of  a  quantum fluid exceeds the
average  interatomic  separation. Then the variables of the continuous
medium  can  only  be  assigned  values, at each mathematical point of
space, in a probabilistic sense.

The  idea  to  describe  superfluid  helium  as a continuous medium at
microscopic  scales  has  been  successfully  used for decades. Atkins
\cite{Atkins}  used  it  in  the  1950s  to  describe  the mobility of
electrons  and  ions  in  $HeI\!I$,  when  he  introduced  bubbles and
snowballs  of  microscopic  size.  Lately  the  vortices  in superfluid helium with cores of sizes of the
scale of interactomic distances are being studied extensively, see for
example  Ref.\cite{Natsik}  and  the  references  cited  there.  Some recent simulations on the dynamics of atoms in helium nanodroplets   \cite{simulations1,simulations2}   also   affirm   that
$HeI\!I$  is  well  described  as  a  continuous medium at microscopic
scales.

As  shown  in \cite{PRB}, application of the methods of theory of continuous
medium  at  microscopic  scales  demands  the relations between the
variables  of  continuous medium to become nonlocal. In the work \cite{Pitaevskii}, the
nonlocal  hydrodynamics  was introduced to describe small oscillations
in  superfluid  helium, and in \cite{P&S,P&S2} it was used to describe
ripplon-roton  hybridization  and dispersion relation of ripplons. The
nonlocality  allows  one  to  analyse  a  continuous  medium  with  an
arbitrary  dispersion  relation.  This  posibility  was  discussed  in
\cite{Whitham}.   However,   the  theoretical  justification  of  this
approach remained on the intuitive level until the work \cite{PRB}.

In  this  paper,  following  \cite{PRB}, the  quasiparticles  are  described as wave packets propagating in the superfluid.  The  long  wavelength  excitations are phonons, while the
short  wavelength  ones  are  rotons.  $R^-$  rotons correspond to the
descending  part  of  the  dispersion  curve  and  have negative group
velocity,  i.e.  they  propagate  in  the  direction opposite to their
momentum. This simple model allows us to use the methods of the theory
of  continuous  medium and avoid the difficulties that appear in other
phenomenological models, such as \cite{P&S,P&S2}.

So,  with  the help of boundary conditions for the continuous media at
the   interface,  we  find  the  creation  probabilities  of  all  the
quasiparticles'   creation  when  any  of  them  is  incident  on  the
interface,  as the energy reflection and transmission coefficients for
the  corresponding  wave  packets.  The method allows us to obtain the
analytical  expressions  for  the probabilities as functions of angles
and frequency.

In  section 2 of the paper we formulate the problem and obtain  the  general
solution of the nonlocal equations of the quantum fluid  in  the
half-space.  We  consider
  a parameterised dispersion  relation  that is a good
approximation to the measured dispersion curve of superfluid helium.
The solution is sought
in  the  form  that  generalizes  the solution for a monotonic
dispersion   of  a  general  form  obtained  in  \cite{JLTP2006}.  The
consequences of using the boundary conditions are discussed, these
include multiple
critical angles, backward refraction and retro-reflection (or
Andreev reflection \cite{Andreev}) of phonons and rotons (see also \cite{P&S2}).

In  section 3 the boundary conditions are used to derive both the amplitude and
energy reflection  and  transmission  coefficients  for any incident wave for
arbitrary  incidence  angles.  The  preliminary  results for  rotons at normal
incidence   were discussed in the authors' report at the
conference \cite{VANT}.

  In section 4 the energy flows  through the
interface due to phonons, $R^-$, and $R^+$ rotons are calculated as
functions of temperature. The contribution of the $R^-$ rotons to the
energy flows, in both directions, are shown to be small. This  means that $R^-$
rotons are hardly
created by a solid heater and are poorly detected by a solid bolometer. This
explains why $R^-$ rotons could not be detected in direct experiments until the
work \cite{exp2}. There they were created by a source made up of
two heaters facing each other which allowed mode changes, and
detection was achieved by quantum
evaporation.

The results obtained in this work can be used in other fields of physics. In
particular they are important for classical acoustics, where the problem of
wave transmission through an interface has been solved only for the case when
the dispersion relations of both adjacent media are strictly linear (see, for
example, \cite{Brehovskih}), let alone non-monotonic. This problem concerning
real media, with nonlinear dispersion, was of interest in the middle
of the last
century \cite{Mandelshtamm} and is still relevant today
\cite{Shelling1,Shelling2}.

\section{Derivation of equations and their solution}

\subsection{Problem Formulation}

Let us consider two continuous media separated by a sharp interface $z\!=\!0$.
In the region $z\!<\!0$ there is an ordinary
continuous medium with sound velocity $s_{sol}$ and equilibrium density
$\rho_{sol}$. For the solid we only take into account longitudinal
waves.

Taking  into  account  the transverse waves can be done in the same
framework  of  theory  of  continuous  medium and does not present any
difficulty.  However,  the  calculations  become much more cumbersome,
while  on  the  whole  the  situation does not change. Due to the very
small  impedance  of the solid-helium interface (see section 2.4 and below), the
reflection  coefficients  hardly  change  at  all. For the transmitted
waves  additional  critical  angles  appear, corresponding to the sound
velocity of the transverse waves. Also it should be noted that taking
into  account  both the longitudinal and transverse waves in the solid
allows  one to consider the contribution of Rayleigh waves, which give
contributions to the transmission coefficients of He II quasiparticles
into  the  solid  at  fixed  incidence angles. For phonons with linear dispersion
this problem was solved in \cite{Rayleigh}.. This problem for the
helium-solid interface may be the subject of next paper.

The region $z\!>\!0$ is filled with the quantum
fluid with equilibrium density $\rho_0$ and dispersion relation $\Omega(k)$
such that
\begin{equation}
     \label{OmegaRot}
     \Omega^{2}(k)=s^{2}k^{2}
     \left\{
     1+2\lambda\frac{k^2}{k_{g}^2}+\frac{k^4}{k_{g}^4}
     \right\}.
\end{equation}

Here $s$ is sound velocity at zero frequency, $k_{g}$ is wave vector that
determines the scale of the curve and $\lambda$ determines the form of the
curve. For a range of parameters, this relation is a good approximation to the
measured non-monotonic dispersion relation of superfluid helium (see Fig.\ref{Fig1}).
For $\lambda\!<\!-1$ there are real $k$ such that $\Omega^{2}(k)\!<\!0$, which
is nonphysical, and for $\lambda\!>\!-\sqrt{3}/2$ the curve is monotonic. For
$\lambda\!\in\!(-1,-\sqrt{3}/2)$ the curve has the roton minimum at
$k\!=\!k_{rot}$ and maxon maximum at $k\!=\!k_{max}$ as it should. We adopt
the following set of values: $s\!=\!230.7$ m/s, $k_g\!=\!1.9828~
\textrm {\AA}^{-1}$,
and $\lambda\!=\!-0.9667$. Then the dispersion curve has the following
parameters: the coordinates of roton minimum
$k_{rot}\!=\!0.9670k_{g}\!=\!1.913~\textrm {\AA}^{-1}$ and
$\Delta_{rot}\!=\!\hbar\Omega(k_{rot})/k_{B}\!=\!8.712$ K ($k_{B}$ is
the Boltzmann
constant); the maxon maximum is
$\Delta_{max}\!=\!\hbar\Omega(k_{max})/k_{B}\!=\!13.8$ K. These values are the
experimentally measured parameters of superfluid helium dispersion curve at
the saturated vapor pressure \cite{neutron}.

  We describe this quantum fluid by nonlocal hydrodynamics as
developed in Ref. \cite{PRB}. Accordingly, the quantum fluid, as well as
the ordinary fluid on the other side of the interface, obeys the linearized
equations of continuous media
\begin{equation}
  \label{ContMed}
  {\displaystyle
\frac{\partial\rho}{\partial t}=-\rho_{0}\nabla \mathbf{v} \quad;\quad
\frac{\partial\mathbf{v}}{\partial t} =-\frac{1}{\rho_{0}}\nabla P,
}
\end{equation}
where $\mathbf{v}$ is hydrodynamic velocity, $\rho$ and $P$ are the deviations
of density and pressure from the respective equilibrium values (for brevity we
refer to them below as just density and pressure). The difference is that
pressure and density in the quantum fluid are related through the non-local
relation
\begin{equation}
\label{nonlocality}
  \rho(\mathbf{r})=
  \int\limits_{z'>0} d^{3}r'
h(|\mathbf{r}-\mathbf{r}'|)P(\mathbf{r}') ,
\end{equation}
in which the integration domain is the region filled by the quantum fluid
\cite{JLTPold}.

The suggested model well describes the interface between superfluid helium and
a solid, because for solids the relationships are local. In the
frequency range of the dispersion curve of HeII,
the dispersion laws of most solids, such as the heater materials of
copper or gold, are
very close to linear and they can be described as ordinary continuous media.

Equations (\ref{nonlocality}) and (\ref{ContMed}) lead to the
integro-differential equation for pressure
\begin{equation}
\label{EQP}
\triangle
P(\mathbf{r},t)=\!\int\limits_{z'>0}\! d^{3}r'\,
h(|\mathbf{r}-\mathbf{r}'|)\ddot{P}(\mathbf{r}',t),
\end{equation}
that is set for $x,y,t\!\in\!(-\infty,\infty), \, z\!\in\!(0,+\infty)$. In the
infinite medium, when the integration and definition domains are infinite, the
Fourier transform of (\ref{EQP}) gives us the relation between the Fourier
transform of the kernel $h(r)$ and the dispersion relation of the fluid
$\Omega(k)$ \cite{PRB}:
\begin{equation}
\label{Hk}
h(k)=\frac{k^2}{\Omega^{2}(k)}.
\end{equation}
For the dispersion relation (\ref{OmegaRot}) we obtain from   the
Fourier transform of Eq. (\ref{Hk}):
\begin{equation}
     \label{Hr}
     h(r)=
     \frac{k_{g}^4}{4\pi s^2 r}
     \frac{1}{k_{+}^{2}-k_{-}^{2}}
     \left(
     e^{ik_{+}r}-e^{-ik_{-}r}
     \right),
\end{equation}
where $k_{+}$ and $(-k_{-})$ are the poles of $h(k)$ in the upper half-plane
$\mathbf{C}_{+}$
\begin{equation}
     \label{K+-}
     k_{\pm}=k_{g}
     \left(\sqrt{1-\lambda}\pm i\sqrt{1+\lambda}\right)/\sqrt{2},\quad
     k_{+}=k_{-}^{\ast}\in\mathbf{C}_{+}.
\end{equation}
Here the asterisk denotes the complex conjugate and $\mathbf{C}_{+}$
is the upper complex half-plane. Due to the last condition, the
kernel, Eq. (\ref{Hr}), despite the complex notation, is real.

There is no convolution product in Eq. (\ref{EQP}), either in the sense of one-
or two-sided Fourier transform or Laplace transform, because the lower limit by
$z'$ is finite while the kernel is symmetrical $h(\mathbf{r})\!=\!h(r)$.

We consider the problem of waves transfering through the interface. As the
equations (\ref{ContMed}) are local and coincide with the  notation
used in the equations
of ordinary ideal continuous medium, the two boundary conditions on the
interface (local!) are obtained from their integral forms, in the
usual way, using
the theory of continuous medium:
  \begin{equation}
\label{boundary}
\left\{
\begin{array}{rl}
     P(x,y,z\!=\!-0,t)&=P(x,y,z\!=\!+0,t),\\
     V_{z}(x,y,z\!=\!-0,t)&=V_{z}(x,y,z\!=\!+0,t).
\end{array}
\right.
\end{equation}
By applying the solutions of the equations of continuous media, on
the both sides
of the interface, the boundary conditions (\ref{boundary}) give us
the solution in the
whole space and thus provide us with all the coefficients of reflection and
transmission. The solution in the solid is well-known, and the solution in the
quantum fluid is derived in the following subsection.

\subsection{Solution of Eq. (\ref{EQP}) in half-space}

The equation that determines the relation between $k$ and $\omega$
  \begin{equation}
     \label{DispEq}
     \Omega^{2}(k)=\omega^{2}
\end{equation}
  with $\Omega^{2}(k)$ from Eq. (\ref{OmegaRot}) is  sixth order with
respect to $k$. Its six roots are functions of $\omega$ and are denoted
$k_{\mu}$ for $\mu\!=\!1,\ldots,6$. We note that if we used more
terms in the polynomial $\Omega^{2}(k)$, then the higher order equation for
$k$ would give 6 real roots and the other roots would be imaginary.

In the problem of waves transfering through the interface, the two boundary
conditions (\ref{boundary}) can be satisfied for all $x$, $y$, and $t$ only if
all of the waves present on both sides of the interface have the same frequency
$\omega$ and tranverse component of wave vector $\mathbf{k}_{\tau}$. A single
monochromatic wave is not a solution of Eq. (\ref{EQP}). Therefore, as there
are in total six roots of Eq. (\ref{DispEq}), we search for the solution as a
sum of six monochromatic waves, with the same frequency $\omega$ and
transverse component of wave vector $k_{\tau}$ (the $y$ axis is chosen along
$\mathbf{k}_{\tau}$), i.e. with the form
\begin{equation}
     \label{Psum6}
     P(\mathbf{r},t)=\sum\limits_{\mu=1}^{6}A_{\mu}
     \exp\left[i(\mathbf{k}_{\mu}\mathbf{r}-\omega t)\right].
\end{equation}
Here the vectors $\mathbf{k}_{\mu}$ are
\begin{eqnarray}
     \label{Kalpha}
     \mathbf{k}_{\mu}&=&k_{\mu z}\mathbf{e}_{z}+k_{\tau}\mathbf{e}_{y},\\
     \label{KalphaSq}
     k_{\mu}^{2}&=&k_{\mu z}^{2}+k_{\tau}^{2}.
\end{eqnarray}
  The transverse component $k_{\tau}$ is real for physical reasons, but
$k_{\mu}$ and $k_{\mu z}$ can be either real or complex, as we have to ensure
the boundedness of our solution only in the half-space $z\!>\!0$.

  After substitution of Eq. (\ref{Psum6}) into Eq. (\ref{EQP}), we obtain the
system of equations for the amplitudes $A_{\mu}$ and the equations
for $k_{\mu z}$, as
functions of $\omega$ and $k_{\tau}$. The system for $A_{\mu}$ is
\begin{equation}
     \label{alpha}
     \left\{
     \begin{array}{c}
         \sum\limits_{\mu=1}^{6}A_{\mu}(k_{\mu z}-k_{+z})^{-1}=0,\\
         \sum\limits_{\mu=1}^{6}A_{\mu}(k_{\mu z}+k_{-z})^{-1}=0,
     \end{array}
     \right.
\end{equation}
where
\begin{equation}
     \label{K+-z}
     k_{\pm z}^2=k_{\pm}^2-k_{\tau}^2,\quad
k_{+z}=k_{-z}^{\ast}\in\mathbf{C}_{+}.
\end{equation}
  The equations for $k_{\mu
z}$ are reduced to the form
\begin{equation}
     \label{DispEqSq}
     \Omega^{2}\left(k_{\mu}^{2}=k_{\mu z}^{2}+k_{\tau}^{2}\right)
     =\omega^{2}\quad\mbox{for}\;\mu\!=\!1,\ldots,6.
\end{equation}

The system of two homogeneous equations for the amplitudes $A_{\mu}$
(\ref{alpha}) ensure that no (nontrivial) solutions exist with less than 3
non-zero amplitudes $A_{\mu}$, i.e. there are no eigensolutions of Eq.
(\ref{EQP}), in the  half-space, consisting of less than 3
monochromatic waves. This
is the consequence of the nonlocality, which changes equation (\ref{EQP}) itself in the
presence of the interface. In infinite space, on the contrary, the domain
of integration is the whole space, and the equation is solved by the
Fourier transform,
and its solution is a superposition of plane waves with dispersion
(\ref{OmegaRot}).

A solution of (\ref{EQP}), with the smallest possible number of waves
being three, is
constructed in the form (\ref{Psum6}), with three terms out of six, by picking
a subset of any three different roots $\{k_{\alpha},k_{\beta},k_{\gamma}\}$ out
of the set of six $\{k_{\mu}\}_{\mu=1,\ldots,6}$. Then it can be rewritten with
the help of Eqs. (\ref{alpha}) in the form that contains a single amplitude:
\begin{align}
     \label{P}
     &P_{\{k_{\alpha},k_{\beta},k_{\gamma}\}}(\mathbf{r},t)
         =P_{\alpha\beta\gamma}^{(0)}\times\nonumber\\
         &\;\times\left\{
         \frac{(k_{\alpha z}-k_{+z})(k_{\alpha z}+k_{-z})}
             {(k_{\alpha z}-k_{\beta z})(k_{\alpha z}-k_{\gamma z})}
         e^{ik_{\alpha z}z}+
         \right.\nonumber\\
        &\quad+
        \frac{(k_{\beta z}-k_{+z})(k_{\beta z}+k_{-z})}
             {(k_{\beta z}-k_{\gamma z})(k_{\beta z}-k_{\alpha z})}
        e^{ik_{\beta z}z}+
         \\
         &\quad+\left.
         \frac{(k_{\gamma z}-k_{+z})(k_{\gamma z}+k_{-z})}
             {(k_{\gamma z}-k_{\alpha z})(k_{\gamma z}-k_{\beta z})}
         e^{ik_{\gamma z}z}
         \right\}
         e^{i(k_{\tau}y-\omega t)},
         \nonumber
\end{align}
where $P_{\alpha\beta\gamma}^{(0)}$ is chosen so that
$P_{\{k_{\alpha},k_{\beta},k_{\gamma}\}}(\mathbf{r}\!=\!0,t\!=\!0)\!=\!P_{\alpha\beta\gamma}^{(0)}$.

The velocity is obtained from Eqs. (\ref{P}) and (\ref{ContMed}):
\begin{align}
     \label{V}
     &\mathbf{v}_{\{k_{\alpha},k_{\beta},k_{\gamma}\}}(\mathbf{r},t)=
         \frac{P_{\alpha\beta\gamma}^{(0)}}
	{\rho_{0}}\times\nonumber\\
          &\;\times\left\{
         \frac{\mathbf{k}_{\alpha}}{\omega}
         \frac{(k_{\alpha z}-k_{+z})(k_{\alpha z}+k_{-z})}
             {(k_{\alpha z}-k_{\beta z})(k_{\alpha z}-k_{\gamma z})}
         e^{ik_{\alpha z}z}+
         \right.\nonumber\\
        &\quad+
        \frac{\mathbf{k}_{\beta}}{\omega}
        \frac{(k_{\beta z}-k_{+z})(k_{\beta z}+k_{-z})}
             {(k_{\beta z}-k_{\gamma z})(k_{\beta z}-k_{\alpha z})}
        e^{ik_{\beta z}z}+
        \\
         &\quad+\left.
         \frac{\mathbf{k}_{\gamma}}{\omega}
         \frac{(k_{\gamma z}-k_{+z})(k_{\gamma z}+k_{-z})}
             {(k_{\gamma z}-k_{\alpha z})(k_{\gamma z}-k_{\beta z})}
         e^{ik_{\gamma z}z}
         \right\}
         e^{i(k_{\tau}y-\omega t)}.
         \nonumber
\end{align}

As the two conditions (\ref{alpha}) restrict the number of free amplitudes in
Eq. (\ref{Psum6}) from six to four, any four linear-independent solutions of
the form (\ref{P}) constitute the basis set of solutions of Eq. (\ref{EQP}) for
given $\omega$ and $k_{\tau}$, and any solution consisting of 4, 5, or 6
monochromatic waves can be represented as their linear combination.

\subsection{Roots of dispersion equation. In- and out-solutions}

The roots of Eq. (\ref{DispEq}) with $\Omega^{2}(k)$ from Eq. (\ref{OmegaRot})
with respect to $k^{2}$ are $k_{i}^{2}\!=\!k_{g}^{2}\xi_{i}$ for $i\!=\!1,2,3$,
where $\xi_{i}$ are the three dimensionless roots of the cubic equation
\begin{equation}
     \label{bicubicEQ}
     \xi^3+2\lambda\xi^2+\xi-\chi^2=0.
\end{equation}
Here $\chi=\omega/(sk_{g})$ is the dimensionless frequency;
$\xi_{i}(\lambda,\chi)$ are
some elaborate complex-valued functions.

The most interesting frequency range is $\chi\!\in\!(\chi_{rot},\chi_{max})$,
where $\chi_{rot}(\lambda)$ and $\chi_{max}(\lambda)$ are the dimensionless
frequencies that correspond to the roton minimum and maxon maximum
respectively.
For such frequencies there are three types of running waves in the quantum
fluid, corresponding to phonons, $R^-$, and $R^+$ rotons. The branches are
numbered in this case in the ascending order of the absolute
values of their wave vectors $k_{i}$:
$0\!<\!k_{1}\!<\!k_{max}\!<\!k_{2}\!<\!k_{rot}\!<\!k_{3}$, so
that $i\!=\!1$ corresponds to phonons, $i\!=\!2$ to $R^-$ rotons, and $i\!=\!3$
to $R^+$ rotons.

We now consider the problem of quasiparticles transfer through the
interface. The quasiparticles are
treated as wave packets that propagate in the two media.
Therefore, when we build the solutions in the quantum fluid, we have to take
into account that wave packets, as well as quasiparticles, propagate with their
group velocities $d\Omega/d\mathbf{k}$ \cite{JML}. So, a wave packet of the
quantum fluid composed of waves with wave vectors close to $\mathbf{k}_{0}$,
with its length $k_{0}<k_{max}$ (so that it is a phonon wave packet) and the
$z$th component $k_{0z}\!>\!0$, propagates away from the interface; but a wave
packet, composed of waves with wave vectors close to one with length
$k_{0}\!\in\!(k_{max},k_{rot})$ (so it is an $R^-$ roton packet) and the $z$th
component $k_{0z}\!>\!0$, propagates towards the interface.

\begin{figure}[ht]
\begin{center}
\includegraphics[viewport=89 435 435 725, width=0.4\textwidth]{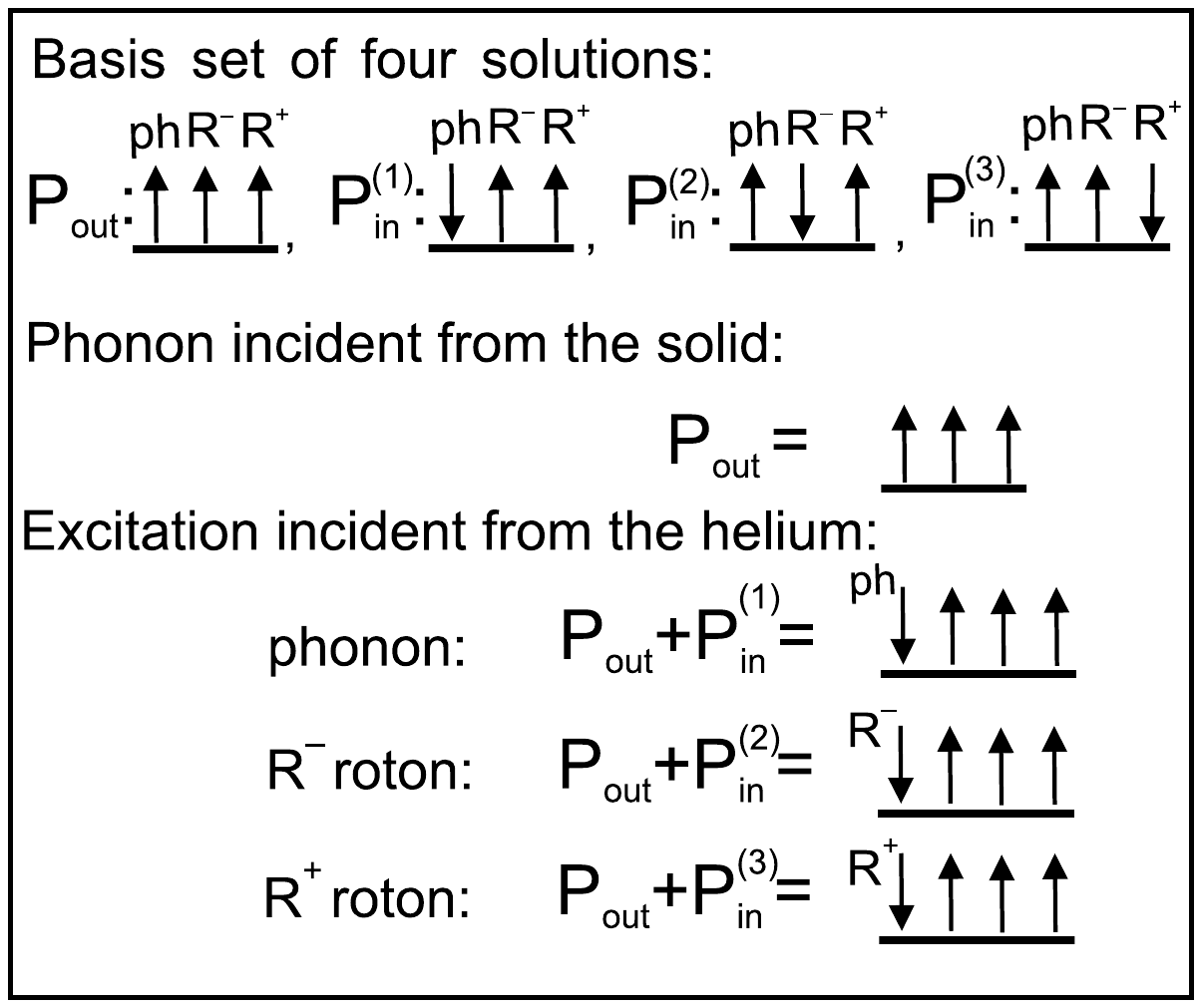}
\parbox{0.45\textwidth}
	{\caption{\label{Fig2} The basis set of solutions in the superfluid helium in the half-space and their sums, which correspond to the different incident excitations.}}
\end{center}
\end{figure}

Let us construct the solution in the quantum fluid $P_{out}$ (the
``out-solution") that is realized when a wave in the solid is incident on the
interface. This solution should contain only such waves that constitute
wave packets traveling away from the interface (i.e. waves with positive group
velocity) or waves that are damped at $z\!\rightarrow\!+\infty$. Picking 3 out
of 6 vectors $\mathbf{k}_{\mu}$ is the same as picking their normal components
$k_{\mu z}$, as they all have the same $k_{\tau}$. The six normal components
$k_{\mu z}$, obtained as solutions of Eq. (\ref{DispEqSq}), are grouped into
three pairs of roots $\pm\sqrt{k_{i}^{2}\!-\!k_{\tau}^{2}}$ for $i\!=\!1,2,3$.
For each pair with the  same $i$, either both roots are real which
occurs for small enough
angles $k_{\tau}\!<\!k_{i}$, or both roots are imaginary for
$k_{\tau}\!>\!k_{i}$. In
the first case one root corresponds to a wave traveling towards the interface,
the other to a wave traveling away from it. The second case, one root
gives a damped wave
in $z\!>\!0$, while the other gives an exponentially unbounded wave.
So, $P_{out}$ contains no more than three waves (and no less because there are
no such solutions) and therefore has the form of Eq. (\ref{P}) (see Fig.\ref{Fig2}).
The squared normal components of the three constituent waves are
\begin{equation}
     \label{KizSquared}
         k_{iz}^{2}\!=\!k_{i}^{2}\!-\!k_{\tau}^{2}\quad\mbox{for}\;i\!=\!1,2,3.
\end{equation}
We define the signs of roots $k_{iz}$ for $P_{out}$ to be made up of waves with
the normal components of wave vectors equal to $k_{1z}$, $k_{2z}$ and $k_{3z}$.
Then taking into account the negative group velocity of $R^-$ rotons
\cite{JML}, for the signs of real $k_{iz}$ we obtain $k_{1z},k_{3z}\!>\!0$ and
$k_{2z}\!<\!0$. If $k_{\tau}\!>\!k_{i}$ for some $i$, the corresponding wave
$\sim\exp{(ik_{iz}z)}$ is bounded in $z\!>\!0$ for $Im\,k_{iz}\!>\!0$. Then in
the general case we have
\begin{align}
     \label{KZsigns}
     k_{\tau}\!\in\!(0,k_{1})&:\;
     0\!<\!k_{1z}\!<\!(-k_{2z})\!<\!k_{3z}\nonumber\\
     k_{\tau}\!\in\!(k_{1},k_{2})&:\;
     0\!<\!(-k_{2z})\!<\!k_{3z},\quad k_{1z}\!\in\!\mathbf{C}_{+}\\
     k_{\tau}\!\in\!(k_{2},k_{3})&:\;
     0\!<\!k_{3z},\quad k_{1z},k_{2z}\!\in\!\mathbf{C}_{+}.\nonumber
\end{align}
  Then the out-solution has the form of Eq. (\ref{P}) with the three wave
vectors picked from the set of six with normal components $k_{1z}$,
$k_{2z}$ and $k_{3z}$:
\begin{equation}
     \label{out-solutions}
     P_{out}=
     P_{\{k_{\alpha},k_{\beta},k_{\gamma}\}}
     \left[k_{\alpha z}\!=\!k_{1z},k_{\beta z}\!=\!k_{2z},k_{\gamma
z}\!=\!k_{3z}\right].
\end{equation}

In order to solve the problem of a wave transfering through the interface from
superfluid helium into the solid, we need also solutions containing waves that
are traveling towards the interface (i.e. wave packets comprised of these waves
should be traveling towards the interface). We define them in the way, which is
illustrated by Fig.\ref{Fig2}. The solution $P_{in}^{(1)}$ is constructed of waves with
$z$th components of wave vectors $(-k_{1z})$, $k_{2z}$ and $k_{3z}$, with the
amplitudes related through Eqs. (\ref{alpha}). Solution $P_{in}^{(2)}$ is
constructed of waves with $k_{1z}$, $(-k_{2z})$ and $k_{3z}$. The last one,
$P_{in}^{(3)}$ contains waves with $k_{1z}$, $k_{2z}$ and $(-k_{3z})$. Then the
three sorts of in-solutions, that correspond to the three types of the incident
waves, can be written in the form
\begin{equation}
     \label{in-solutions}
      P_{in}^{(i)}=\left.P_{out}\right|_{k_{iz}\rightarrow(-k_{iz})}\quad
      \mbox{for}\;
      i\!=\!1,2,3.
\end{equation}
The $P_{in}^{(i)}$ solution corresponds to the incident wave of type $i$. So, a
linear combination of $P_{out}$ and $P_{in}^{(2)}$ consists of one $R^-$ roton
wave ($i\!=\!2$) that corresponds to the $R^-$ roton wave packet incident on
the interface, and all three waves that correspond to the reflected phonon,
$R^-$, and $R^+$ roton wave packets. The amplitudes of the phonon and $R^+$
roton waves are the sums of the amplitudes of those waves present in both
$P_{out}$ and $P_{in}^{(2)}$. This is the solution in $z\!>\!0$ realized when
an $R^-$ roton is incident on the interface. The four solutions
(\ref{out-solutions}) and (\ref{in-solutions}) are linearly-independent due to
their structure and can be used as the basis set of solutions as mentioned at
the end of the previous subsection.

For $\chi\!\in\!(\chi_{rot},\chi_{max})$ the roots $k_{iz}$ are fully defined
by Eqs. (\ref{KizSquared}) and (\ref{KZsigns}). For $\chi\!\in\!(0,\chi_{rot})$
the roots $\xi_{2,3}$ are complex and $\xi_{2}\!=\!\xi_{3}^{\ast}$; then
$k_{2z}$ and $k_{3z}$ are defined so that the roton waves
$\sim\exp{(ik_{2z,3z}z)}$ are damped, so
$k_{2z}\!=\!-k_{3z}^{\ast}\!\in\!\mathbf{C}_{+}$. In the limit
$\chi\!\rightarrow\!0$ the phonon waves have almost linear dispersion
$k_{1}\!\approx\!\omega/s$, and it can be shown that
$k_{2z}\!\rightarrow\!k_{+z}$ and $k_{3z}\!\rightarrow\!-k_{-z}$. Therefore the
amplitudes of all roton waves, that contain multipliers $(k_{2z}-k_{+z})$ and
$(k_{3z}+k_{-z})$, go to zero, see (\ref{P}), and the general
solution tends to the ordinary
superposition of incident and reflected phonon waves, with wavelengths much
greater than the scale of nonlocality $|k_{\pm}|^{-1}$. This limiting case can
also be obtained by passing to the longwave limit
$h(r)\!\rightarrow\!\delta(r)/s^{2}$ in Eq. (\ref{EQP}), thus making
it a local wave equation.

The equation (\ref{EQP}) was solved earlier in work \cite{JLTP2006} for the
case of arbitrary but monotonic dispersion relation. The Wiener and Hopf method
was used there, the application to this type of problem was
suggested in \cite{JLTPold} and developed in \cite{PhNT}. It is much more
general and seems to be more rigorous than the method used here, though less
straight-forward. The solution of \cite{JLTP2006} can be generalized to the
nonmonotonic case and can be shown to yield for the dispersion relation
(\ref{OmegaRot}) exactly the solutions (\ref{out-solutions}) and
(\ref{in-solutions}).

In  the  work  \cite{P&S}  the  problem  analogous  to (\ref{EQP}) was
considered  in  order  to  investigate the hybridization of rotons and
ripplons.  There  the problem in half-space was replaced by the one in
the  infinite  medium  with symmetrical extrapolation of the solutions
with  respect  to  variable  $z$.  However,  in  contrast  to  the usual
differential  (local)  equations, for an integral equation such as (\ref{EQP})
the solution cannot be thus extrapolated symmetrically to z<0. Indeed,
if  we  consider  formally the solution of Eq. (\ref{EQP}) in $z\!<\!0$, it would be
defined  unambiguously by the solution in $z\!>\!0$ (from Eq. (\ref{P})), through
the  integral  of  Eq.  (\ref{EQP}).  Direct  substitution shows that the full
solution  is  not  even.  This  might  be  the  reason that the method
\cite{P&S} gave wrong results near the surface waves threshold and was
then  rejected;  in  the next paper by the authors \cite{P&S2} another
approach was used for that problem.

\subsection{Multiple critical angles and Andreev reflection}

Even before applying the boundary conditions (\ref{boundary}) and finding the
solution in the whole space, we can use the fact that two linear boundary
conditions for the variables of continuous media are satisfied on the interface
and derive a number of important consequences. First of all, we see that the
solution is always constructed in a such a way that there are in total four
outgoing (i.e. reflected and transmitted) waves; one in the solid and three
in the quantum fluid. This is because the four conditions on the waves
amplitudes, two from Eq. (\ref{boundary}) and two from Eq. (\ref{alpha}), can
be all satisfied only when there are at least four outgoing waves. On the other
hand, they can be at most four because, either in the  formulation of
the problem there
is only one incident wave, or the requirement that the  solution is
bounded when some
of $k_{iz}$ are complex.

\begin{figure}[t]
\begin{center}
\includegraphics[viewport=130 480 435 725, width=0.4\textwidth]{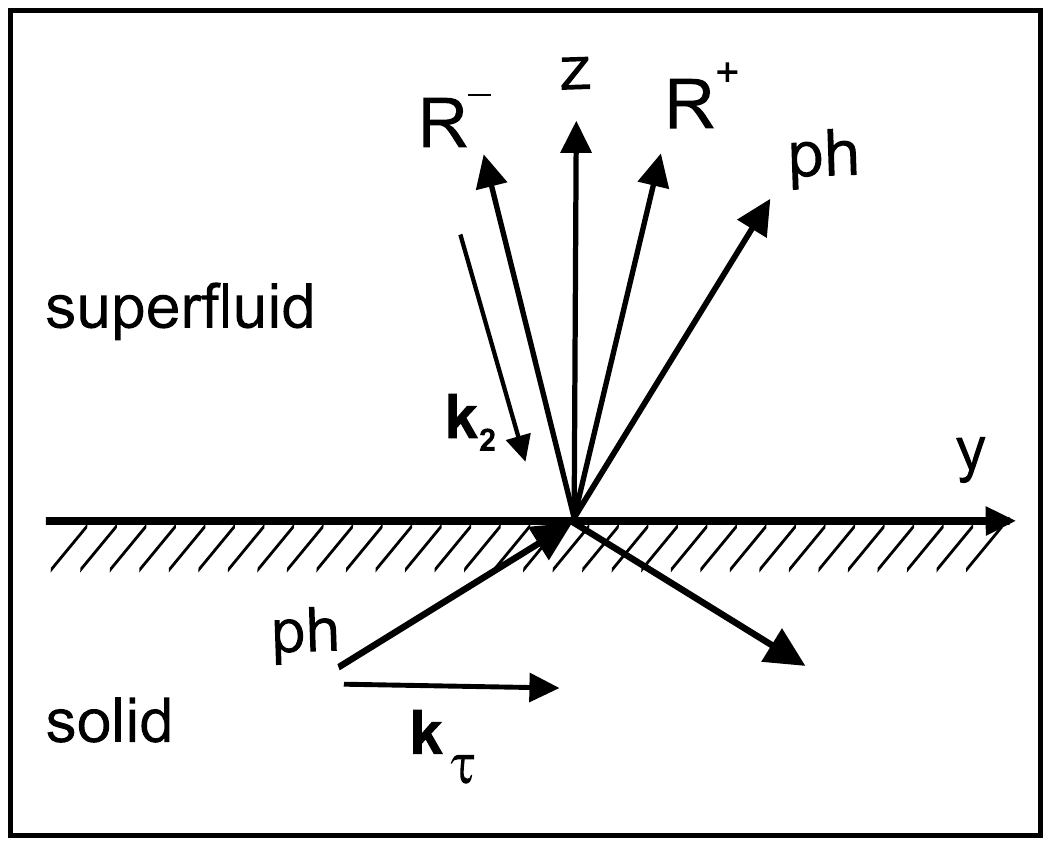}
\parbox{0.45\textwidth}
	{\caption{\label{Fig3} When a phonon in the solid is incident on the interface, three quasiparticles, a phonon, $R^-$ roton, and $R^+$ roton, are created in superfluid helium with the same $\omega$ and $k_{\tau}$. The created $R^-$ roton propagates backward in the transverse direction (i.e. retro-refracted).}}
\end{center}
\end{figure}

Furthermore, the two boundary conditions (\ref{boundary}) imply that all of the
waves constituting the full solution have the same frequency $\omega$ and
tangential component of wave vector $k_{\tau}$. When wave $i$ is propagating at
angle $\theta_{i}$ to the normal to the interface,
$k_{\tau}\!=\!k_{i}\sin{\theta_{i}}$. Then if one of the waves is incident, the
corresponding angle and $k_{\tau}$ are set and all the other transmission and
reflection angles are determined by the generalization of Snell's law
\begin{equation}
\label{Snell}
     \frac{\sin{\theta_{sol}}}{s_{sol}}=
     \frac{\sin{\theta_{1}}}{s_{1}}=
     \frac{\sin{\theta_{2}}}{s_{2}}=
     \frac{\sin{\theta_{3}}}{s_{3}}.
\end{equation}
Here $s_{i}\!=\!\omega/k_{i}(\omega)$ are the phase velocities of the
corresponding waves, that depend on frequency;
$\mathbf{k}_{sol}\!=\!k_{sol\,z}\mathbf{e}_{z}+k_{\tau}\mathbf{e}_{y}$ is the
wave vector of the wave in the solid and
$s_{sol}\!=\!\omega/k_{sol}(\omega)\!=\!const$. The reflection angle for the
wave of the same type as the incident one is equal to the incidence angle.

 From now on we will consider $s_{sol}\!>\!s$, as is the case when superfluid
helium is adjacent to a solid. Usually even the strong inequality holds. Then
we have
\begin{equation}
     \label{soundvelocities}
     s_{sol}>s_{1}>s_{2}>s_{3}>0.
\end{equation}
If a wave from the solid is incident at $\theta_{sol}$, it is reflected at the
same angle and the three waves are transferred into helium at angles
$\theta_{i}\!<\!\theta_{sol}$. The $R^-$ roton wave, as opposed to the others,
due to its negative group velocity, propagates backward in the tangential
direction (i.e. in the direction $y\!\rightarrow\!-\infty$, see Fig.\ref{Fig3}).

Assume a wave $i$ is incident from helium at $\theta_{i}$. Then according to
Eq. (\ref{Snell}) the transmitted wave in the solid has
$\sin{\theta_{sol}}\!<\!1$ for $\sin\theta_{i}\!<\!s_{i}/s_{sol}$, and if the
incidence angle is greater than the critical value, $k_{sol\, z}$ is imaginary
and the wave in the solid is exponentially damped. Thus we obtain the three
angles of full internal reflection
\begin{equation}
     \label{CritAnglesSolid}
     \sin{\theta_{i}^{cr}}=s_{i}/s_{sol}\quad\mbox{for}\;i\!=\!1,2,3.
\end{equation}
In the same way there are three new critical angles defined for $i\!>\!j$ (so
that $s_{i}\!<\!s_{j}$):
\begin{equation}
     \label{CritAnglesHe}
     \sin{\theta_{ij}^{cr}}=s_{i}/s_{j}\!<\!1\quad\mbox{for}\;\{i,j\}=\{2,1\},\{
3,2\},\{3,1\}.
\end{equation}
If a wave $i$ is incident and $\theta_{i}\!<\!\theta_{ij}^{cr}$, then wave $j$
(with $j\!<\!i$) has $\theta_{j}\!\in\!(\theta_{i},\pi/2)$ and
$k_{jz}\!\in\!\mathbf{R}$. For $\theta_{i}\!>\!\theta_{ij}^{cr}$ $\,
k_{jz}^{2}\!<\!0$, the $j$-th wave is damped and the corresponding
quasiparticle is not created.

In an ordinary fluid, the group velocity of a wave packet is the same
as the sound
velocity and is constant. When a wave is incident on the interface the
reflected wave has the same wave number and due to preservation of the
transverse component of the wave vector $\mathbf{k}_{\tau}$, it is reflected
forward. This qualitative picture is maintained in the majority of all known
physical systems. However, when the reflected wave (or quasiparticle) is
qualitatively different from the incident, another possibility can be
realized. So, when an electron of a normal metal is incident on the interface
with a superconductor, it can be retro-reflected and converted into the hole of
negative effective mass that travels back along the same line as the incident
electron. This effect was discovered by Andreev (see ref. \cite{Andreev}) and
is called Andreev reflection or retro-reflection.

In our case when a helium quasiparticle is incident on the interface,
three different
quasiparticles are created in helium, with corresponding probabilities. If for
example an $R^-$ roton is created on the interface, along with other
quasiparticles, when a phonon is incident, for the sake of brevity we will
refer to this as ``the phonon is reflected into $R^-$ roton". While phonons and
$R^+$ rotons behave in these processes like ordinary quasiparticles, $R^-$
rotons propagate in the direction opposite to their wave vector, due to the
negative group velocity as already mentioned above. Therefore when a phonon or
$R^+$ roton is incident on the interface, the phonons and $R^+$ rotons are
reflected forward, while $R^-$ rotons are reflected backward, or
\textit{retro-reflected}. In this way the transverse components of wave
vectors $\mathbf{k}_{\tau}$ are all equal. Likewise, when an $R^-$ roton is
incident, the phonon and $R^+$ roton are retro-reflected. Thus we have
described the effect of Andreev reflection of helium phonons and rotons.

\begin{figure}[b]
\begin{center}
\includegraphics*[viewport=132 530 435 754, width=0.4\textwidth]{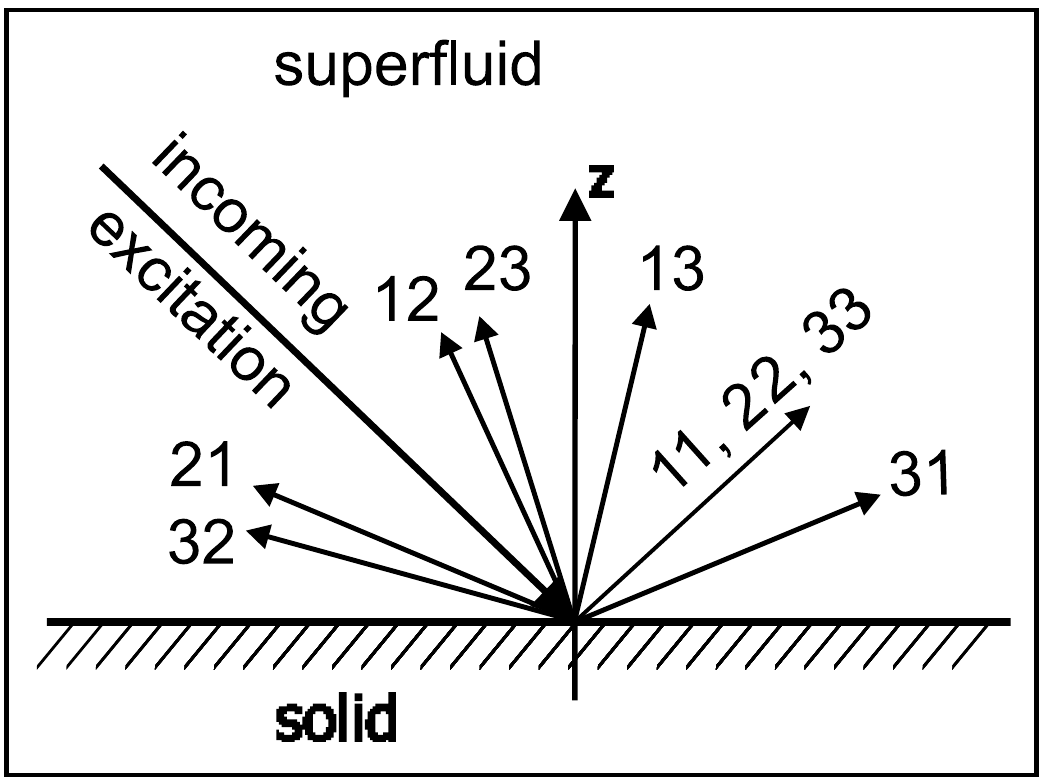}
\parbox{0.45\textwidth}
	{\caption{\label{Fig4} When a single beam of quasiparticles is incident on the interface, a set of reflected beams is created. The beam marked $ij$ consists of quasiparticles of type $j$ created by incident quasiparticles of type $i$. The beams $ii$ are reflected specularly, while all others propagate in different directions. The beams $12$, $21$, $23$, and $32$ are retro-reflected.}}
\end{center}
\end{figure}

If a monochromatic beam of phonons and rotons is incident on the interface at
some angle $\Theta$, there will be up to seven reflected beams (see Fig.\ref{Fig4}). We
denote them as $ij$, which  means ``the beam of quasiparticles of type
$j$ created on the interface by the incident quasiparticles of type $i$". The
beams $11$, $22$ and $33$ are reflected forward at the incidence
angle $\theta$, and
therefore constitute a single beam $ii$. Beams $13$ and $31$ are also reflected
forward, while $12$, $21$, $32$ and $23$ are reflected \textit{backward}. Due
to the relations (\ref{Snell}), the beams $32$ and $21$ are reflected at angles
greater than $\theta$, and beams $12$ and $23$ at angles less than $\theta$.

The reflection angles depend on $s_i$, which are functions of frequency.
Therefore with an incident beam which is  non-monochromatic, the
reflected beams all
become angularly diffused \cite{JLTP2006}, except for the beam $ii$.
However, the relations (\ref{soundvelocities}) hold at all frequencies, and
therefore the qualitative picture is not modified. If we place a detector on
the same side from the normal as the source at greater angles, it should
register the $R^-$ rotons of beam $32$ and phonons of beam $21$, which were
\textit{retro-reflected}. Likewise the detector at smaller angles should
register the $R^-$ rotons of beam $12$ and $R^+$ rotons of beam $23$. Such an
experiment could be carried out in order to verify qualitatively the current
theory.

A successful experiment would very much depend  on the intensities of
the beams to be detected, and therefore on the different
creation probabilities for the quasiparticles at the interface.
The derivation of these probabilities is the subject of the
next section.

\section{Reflection and transmission coefficients}

\subsection{Phonon in the solid  incident on the interface}

  When a phonon in the solid is incident on the interface, it is reflected and
  three quasiparticles of different types are created in
helium which travel away from
the interface (see Fig.\ref{Fig3}). The probability of quasiparticle creation
is the fraction of incident
energy which is reflected or transmitted  as the corresponding wave
packet. Amplitude reflection and transmission coefficients can be derived in
the approximation of plane waves \cite{PhNT}.

In this approximation we consider a plane wave with frequency $\omega$ and wave
number $k_{sol}(\omega)\!=\!\omega/s_{sol}$ incident on the interface at angle
$\theta_{sol}$ to the normal. Then the solution in the solid is the sum of
the incident and reflected waves, the solution in the quantum fluid
is $P_{out}$
from Eq. (\ref{out-solutions}), which consists of three waves. All the waves
have the same frequency $\omega$ and transverse component of wave vector
$k_{\tau}\!=\!k_{sol}\cos\Theta_{sol}$. The pressure amplitude of each wave
$P_{i}$ is the full coefficient multiplying the exponent
$\exp(ik_{iz}z)$ in the
out-solution. With the help of the boundary conditions (\ref{boundary}), the
amplitudes of all the waves are expressed through the amplitude of the incident
wave. Then after some transformations, the amplitude reflection coefficient,
defined as the ratio of pressure amplitudes in the reflected and incident
waves, can be expressed in the form
  \begin{equation}
     \label{r->}
     r_{\rightarrow}=\frac
     {f_{z}-Z-i\tilde{\delta}}
     {f_{z}+Z-i\tilde{\delta}}.
  \end{equation}
Here the following notations are used. $Z$ is a real generalization of
impedance
\begin{equation}
     \label{Z}
     Z=Z_{g}\cos{\theta_{sol}},\quad Z_{g}=Z_{0}\chi,
\end{equation}
where $ Z_{0}=(\rho_{0}s)/(\rho_{sol}s_{sol})$ is the ordinary impedance of the
interface at zero frequency; $\chi\!=\!\omega/sk_{g}$ is the dimensionless
frequency as introduced in Eq. (\ref{bicubicEQ}); $\tilde{\delta}$ is a
dimensionless constant
\begin{equation}
     \label{delta}
     \tilde{\delta}(\lambda)=\frac{k_{+z}-k_{-z}}{ik_g}\in\mathbf{R},
\end{equation}
which is real due to Eq. (\ref{K+-z});
\begin{align}
     \label{f}
     &f_{z}=f_{3z}/(k_{g}f_{2z}),
	\quad\mbox{where}\nonumber\\
    & f_{n\,z}=\!
     k_{1z}^{n}(k_{2z}\!-\!k_{3z})
	+k_{2z}^{n}(k_{3z}\!-\!k_{1z})
	+k_{3z}^{n}(k_{1z}\!-\!k_{2z})\nonumber\\
    & \;\,\text{for}\; n\!=\!2,3.
\end{align}
As $s_{sol}\!>\!s_{i}$, the transmission angles for all the waves $\theta_{i}$
are less then $\theta_{sol}$, so $k_{iz}\!\in\!\mathbf{R}$ and $f_{z}$ is a
dimensionless real function of $\chi$ and $k_{\tau}/k_{g}$.

The full transmission coefficient is
$t_{\rightarrow}\!=\!1\!+\!r_{\rightarrow}$; the partial transmission
coefficients $t_{i}^{\rightarrow}$ are defined as the ratios of pressure
amplitudes of each of the three waves in helium $P_{i}$ to the pressure
amplitude of the incident wave. They are obtained in the same way as
$r_{\rightarrow}$:
\begin{eqnarray}
     \label{ti}
     &&t_{i}^{\rightarrow}=t_{\rightarrow}
     \frac{\psi_{i}}{(k_{iz}-k_{jz})(k_{iz}-k_{kz})},\\
    &&\mbox{where}\quad
     \psi_{i}=(k_{iz}-k_{+z})(k_{iz}+k_{-z}) \nonumber
\end{eqnarray}
and the subscripts take values $\{i,j,k\}\!=\!\{1,2,3\}\!+\!perm.$ ($perm.$ is
for permutations). Henceforth the subscripts $\{i,j,k\}$ in the expressions of
the kind of Eq. (\ref{ti}) take the same set of values, unless stated
otherwise.

All the amplitude coefficients are complex-valued functions of frequency and
incidence angle. Therefore there are always nontrivial phase shifts between the
incident, reflected, and transmitted waves.

The energy reflection and transmission coefficients are the
normal components of energy density flux, expressed as  fractions of
the incident energy flux, that are
reflected or transmitted into helium. The energy density flux, in a wave
packet in the quantum fluid, as shown in \cite{JML}, equals the average energy
density multiplied by the group velocity. It was shown in \cite{PhNT}, that the
average energy density in a wave packet or plane wave in the quantum fluid with
velocity amplitude $V_{i}$ is given by the same relation as in the ordinary
liquid, $\rho_{0}|V_{i}|^2$, and from Eq. (\ref{V})
$V_{i}\!=\!P_{i}/(\rho_{0}s_{i})$. The group velocity of wave $i$ can be
obtained from Eq. (\ref{OmegaRot}):
\begin{equation}
     \label{U}
     \left|u_{i}\right|=
     \frac{s^2}{k_{g}^{4}}
     \frac{k_{i}}{\omega}
     \left|
     (k_{i}^{2}-k_{j}^{2})(k_{i}^{2}-k_{k}^{2})
     \right|.
\end{equation}
Taking all of this into account and with the help of Eqs. (\ref{ti}),
after some
transformations we obtain the fractions of the normal component of the incident
wave packet's energy flux that are carried  by waves of each type
$i\!=\!1,2,3$ in helium. Those are the partial energy transmission coefficients
\begin{equation}
     \label{D->i}
     D_{i}^{\rightarrow}=
     \frac{4Z}{(Z+f_{z})^{2}+{\tilde{\delta}}^2}\cdot
     \frac{k_{iz}}{k_{g}}\cdot
     \frac{(k_{iz}+k_{jz})(k_{iz}+k_{kz})}{(k_{iz}-k_{jz})(k_{iz}-k_{kz})}.
\end{equation}
We note that $D_{i}^{\rightarrow} >0$ for all $i=$ 1, 2, 3.
The full energy transmission coefficient can be expressed in the form
\begin{equation}
     \label{D->}
     D_{\rightarrow}=\sum\limits_{i=1}^{3}D_{i}^{\rightarrow}=
     \frac{4Zf_{z}}{(Z+f_{z})^{2}+{\tilde{\delta}}^2}.
\end{equation}

\begin{figure}[t]
\begin{center}
\includegraphics[viewport=93 0 505 300, width=0.45\textwidth]{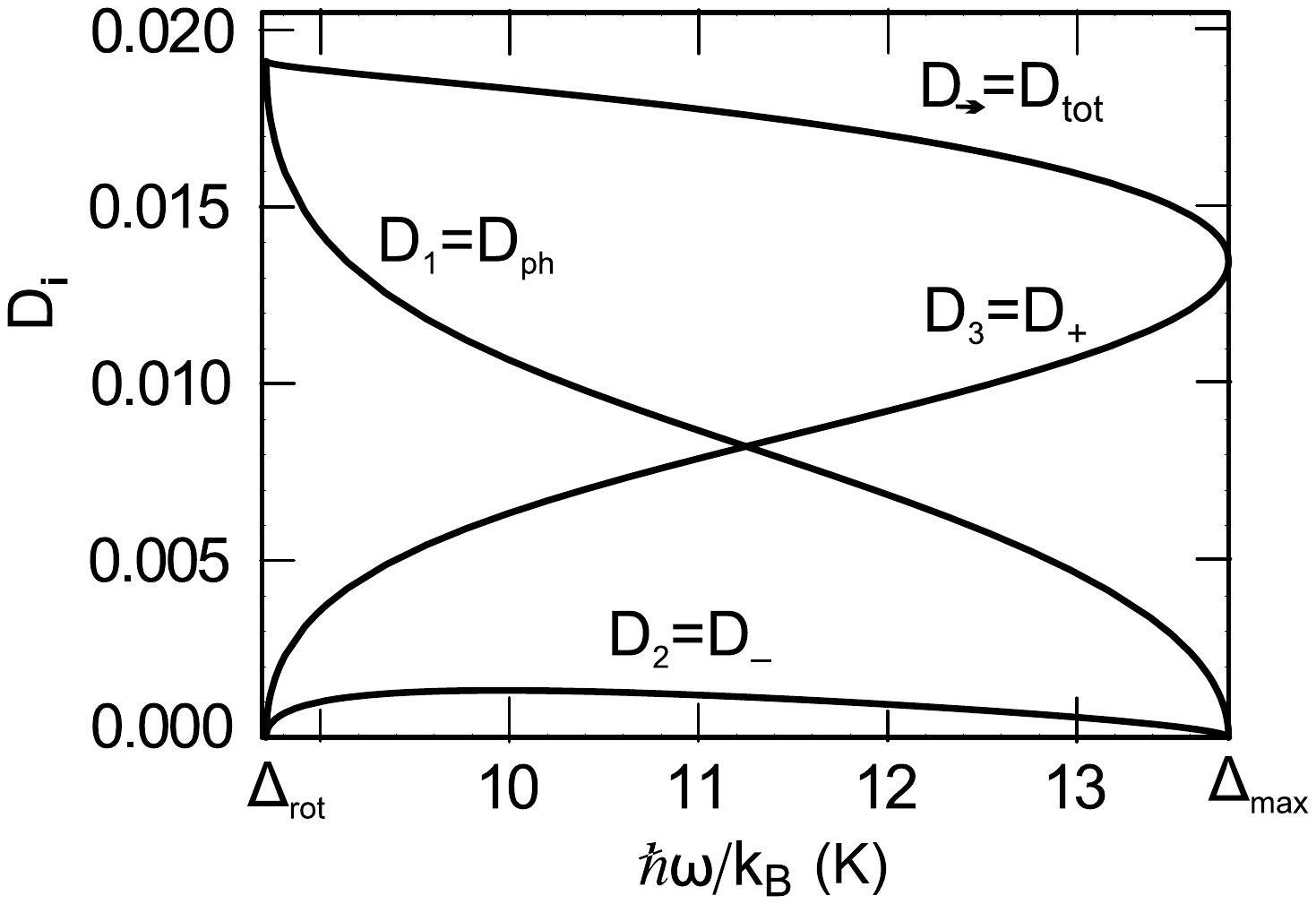}
\parbox{0.45\textwidth}
	{\caption{\label{Fig5} Energy dependence of the transmission coefficients at $\theta\!=\!0$,
with energy in temperature units, for the parameterised dispersion curve (see Fig.\ref{Fig1}) and $Z_{0}\!=\!0.01$. The $R^-$ roton creation probability, $D_2$, is small.}}
\end{center}
\end{figure}

The energy reflection coefficient is $R_{\rightarrow}\!=\!|r_{\rightarrow}|^2$.
Then from Eqs. (\ref{r->}) and (\ref{D->}) after some algebraic transformations
we can show explicitly that
\begin{equation}
     \label{EnConserve}
     R_{\rightarrow}+D_{\rightarrow}=1,
\end{equation}
and so  energy is
conserved when waves go through the interface.
  This also verifies that  the coefficients
$D_{i}^{\rightarrow}$ and $R_{\rightarrow}$ are the probabilities of
the creation of the  corresponding quasiparticles at the interface.

Superfluid helium has a very small density and sound velocity, such that at the
interface with a solid, the strong inequalities $s\!\ll\!s_{sol}$ and
$\rho_{0}\!\ll\!\rho_{sol}$ hold. Then, taking into account Eq.
(\ref{soundvelocities}), we have a set of small parameters
\begin{equation}
     \label{smallness}
     Z_{0}\ll 1,\quad
     s_{i}/s_{sol}\ll 1.
\end{equation}
It can be shown that, due to the first condition of Eq. (\ref{smallness}), in
the sums of Eqs. (\ref{r->}), (\ref{D->}) and (\ref{D->i}) the quantity $Z$ can
be neglected in comparison with the other terms. The second condition of Eq.
(\ref{smallness}) implies that, due to Eqs. (\ref{Snell}), all the transmission
angles into the helium are very small, as is indeed well-known for
the interfaces between
$HeI\!I$ and solids. Then $k_{\tau}^{2}\!\ll\!k_{i}^{2},k_{\pm}^{2}$ and due to
Eqs. (\ref{K+-}) and (\ref{K+-z}) we have $k_{\pm z}\!\approx\!k_{\pm}$. In
this approximation we obtain  $f_{z}\!\approx\!f\!\equiv
\!f_{z}(\chi,\Theta_{sol}\!=\!0)$ and $\tilde{\delta}\!\approx\!\delta\!\equiv
(k_{+}-k_{-})/ik_{g}$. Then the dependance of $D_{i}^{\rightarrow}$
on the incidence angle, from
(\ref{D->i}),is factorized out and is reduced to the
multiplier $\sim\cos{\theta_{sol}}$:
\begin{align}
     D_{i}^{\rightarrow}(\chi,\theta_{sol})&\approx
     \frac{4Z_{g}\cos{\theta_{sol}}}{f^{2}+\delta^2}
     \times\nonumber\\ &\times
     \left.\left(
     \frac{k_{iz}}{k_{g}}\,
     \frac{(k_{iz}+k_{jz})(k_{iz}+k_{kz})}{(k_{iz}-k_{jz})(k_{iz}-k_{kz})}
     \right)\right|_{\theta_{sol}=0}.
     \label{D->iAppr}
\end{align}
  The frequency dependence of the transmission factors, at normal incidence
$\theta_{sol}\!=\!0$, is shown in Fig.\ref{Fig5}. The relative creation
probabilities of
phonons, $R^-$, and $R^+$ rotons are determined by the multipliers in
parenthesis in (\ref{D->iAppr}). They can be rewritten in terms of $k_{i}$,
while taking care of the signs: $k_{1z}$ and $k_{3z}$ at $\theta_{sol}\!=\!0$
are equal to $k_{1}$ and $k_{3}$, but $k_{2z}$ at $\theta_{sol}\!=\!0$ is equal
to $(-k_{2})$ (because of the negative group velocity of $R^-$ rotons). Then
for $i\!=\!2$ we obtain
  \begin{equation}
     \label{D2appr}
     D_{2}^{\rightarrow}\propto
     \frac{k_{2}-k_{1}}{k_{2}+k_{1}}\cdot
      \frac{k_{3}-k_{2}}{k_{3}+k_{2}}\cdot
     k_{2}.
\end{equation}
Both the first and second multipliers here are less than unity. In the
analogous expressions for $D_{1,3}^{\rightarrow}$ one of the two corresponding
multipliers is reversed. So, for the ratio
$D_{2}^{\rightarrow}/D_{1,3}^{\rightarrow}$, the effect is squared
and we obtain
\begin{equation}
     \label{Drot}
     D_{2}^{\rightarrow}\ll D_{1,3}^{\rightarrow}.
\end{equation}
Near the roton minimum, when $\chi\!\rightarrow\!\chi_{rot}$, the $R^-$ and
$R^+$ roton branches merge, their group velocities tend to zero, and so do
their creation probabilities $D_{2,3}\!\rightarrow\!0$ (because they are
proportional to the energy density fluxes, which are proportional to the group
velocities). In Eq. (\ref{D2appr}) the multiplier $(k_{3}\!-\!k_{2})$ comes
from the group velocity.
In the same way
$D_{1,2}^{\rightarrow}\!\rightarrow\!0$ near the maxon maximum
$\chi\!\rightarrow\!\chi_{max}$, where the phonon and $R^-$ roton branches
merge. Thus $D_{2}^{\rightarrow}$ becomes zero at both ends of the frequency
interval in which $D_{2}^{\rightarrow}$ is defined. Both the strong
inequality (\ref{Drot}) and
asymptotic behavior of $D_{2}$ at $\chi\!\rightarrow\!\chi_{rot},\chi_{max}$
(see Fig.\ref{Fig5}) are the consequences of the simple relations
$0\!<\!k_{1z}\!<\!(-k_{2z})\!<\!k_{3z}$ from Eqs. (\ref{KZsigns}), which
reflect the \textit{qualitative} behavior of the  dispersion
curve for superfluid helium, as shown on Fig.\ref{Fig1}.

  The creation probability of $R^-$ rotons at the interface is very
small for all energies.
It should also be noted that, at low temperatures,
the main contribution to the energy flow through the interface is due to
phonons of energies less than the roton gap (i.e. with $\chi\!<\!\chi_{rot}$),
which are not yet taken into account. Thus twe have a convincing explanation
  why $R^-$ rotons were not detected in experiments which created beams
of quasiparticles in helium, by a solid heater, as for example \cite{exp1}.

The expressions for $D_{i}^{\rightarrow}$ (\ref{D->i}), (\ref{D->iAppr}), and
$D_{\rightarrow}$  (\ref{D->}) are written as functions of the incidence angle
or $k_{\tau}$. What can be measured experimentally are the energy flows as
functions of transmission angles. If phonons be incident on the interface
isotropically, then as the transmission angle for each wave is defined by Eqs.
(\ref{Snell}), the quasiparticles of each type are transmitted in a narrow cone
with the cone angle twice the $\theta_{i}^{cr}$. Thus the phonons are injected
into the helium in the widest cone and $R^+$ rotons in the narrowest cone.
For the total transmission coefficient as function of transmission angle
$\theta$ we obtain
\begin{equation}
     \label{Dtheta}
     \begin{array}{l}
     D_{\rightarrow}(\chi,\Theta)\!=\!
     \sum\limits_{i=1}^{3}
     D_{i}^{\rightarrow}(\chi,k_{\tau i}\!=\!k_{i}\sin{\theta}),\\
     \mbox{where}\;D_{i}^{\rightarrow}(\theta)\!=\!0
     \quad\mbox{for}\;\theta\!>\!\theta_{i}^{cr}
     \end{array}
\end{equation}
and $k_{\tau i}$ are the transverse components of the wave vectors  of wave $i$
transmitted at angle $\theta$.

\subsection{Phonon or roton in the helium incident on the interface}

Now let us consider one of the quasiparticles of superfluid helium, phonon or
roton, incident on the interface. In terms of plane waves, a wave $i$ of
frequency $\omega$ and wave vector of length $k_{i}(\omega)$ is incident. The
solution in the solid consists of only one transmitted wave, and solution in
the quantum fluid consists of one incident wave $i$ and three reflected waves
$j\!=\!1,2,3$; it can be represented as a sum of solutions
$P_{out}$ and $P_{in}^{(i)}$ from Eqs. (\ref{out-solutions}) and
(\ref{in-solutions}) (see Fig.\ref{Fig2}). The boundary conditions (\ref{boundary})
enable us to express all the amplitudes through the amplitude of the incident
wave, and thus to obtain the nine amplitude reflection coefficients $r_{ij}$.
The coefficient $r_{ij}$ is the ratio of the pressure amplitudes of
the reflected
wave $j$ to the incident wave $i$ for $i,j\!=\!1,2,3$:
\begin{eqnarray}
     \label{rii}
     r_{ii}&=&-
     \frac{\psi_{i}}{\psi_{i}^{\ast}}\cdot
     \frac{f_{-2z}^{(i)}}{f_{2z}}\cdot
     \frac
     {f_{-z}^{(i)}+Z-i\tilde\delta}{f_{z}+Z-i\tilde\delta};\\
     \label{rij}
     r_{ij}&=&2
     \frac{\psi_{j}}{\psi_{i}^{\ast}}\cdot
     \frac{k_{iz}(k_{i}^{2}-k_{k}^{2})}{f_{2z}}\cdot
     \frac{k_{kz}/k_{g}+Z-i\tilde\delta}{f_{z}+Z-i\tilde\delta}\cdot
     \varepsilon_{ijk}.
\end{eqnarray}
Here the subscripts take values $\{i,j,k\}\!=\!\{1,2,3\}\!+\!perm.$;
$\varepsilon_{ijk}$ is the Levi-Civita symbol, equal to $1$ if
$\{i,j,k\}\!=\!\{1,2,3\},\{2,3,1\},\mbox{or}\,\{3,1,2\}$ and to $(-1)$ if
$\{i,j,k\}\!=\!\{2,1,3\},\{1,3,2\},\mbox{or}\,\{3,2,1\}$;
$Z\!=\!-\!Z_{g}k_{sol\,z}/k_{sol}$ is the generalization of definition
(\ref{Z}). For the incidence angles less than critical
$k_{\tau}^{2}\!<\!k_{sol}^{2}$, $k_{sol\,z}\!<\!0$ and $Z$ is given by Eq.
(\ref{Z}). For greater incidence angles the new notation must be used, as
$\cos\theta$ is not defined. Then $k_{sol\,z}\!\in\!\mathbf{C_{-}}$  for the
wave to be damped in $z\!<\!0$ and therefore $Z\!=\!i|Z|$. The constructions
$f_{-nz}^{(i)}$ are
\begin{align}
     \label{f-}
     &f_{-z}^{(i)}=f_{-3z}^{(i)}/(k_{g}f_{-2z}^{(i)}),\\
     &f_{-nz}^{(i)}=
     \left.f_{nz}[k_{1z},k_{2z},k_{3z}]\right|_{k_{iz}\rightarrow(-k_{iz})}\nonumber\\
     &\quad\text{for}\;i\!=\!1,2,3;\;n\!=\!2,3.\nonumber
\end{align}

The amplitude coefficient of transmission $t_{i}^{\leftarrow}$ for the
incident wave of type $i$ is
\begin{equation}
     \label{ti<-}
     t_{i}^{\leftarrow}=
     \frac{(k_{iz}+k_{jz})(k_{iz}+k_{kz})}{\psi_{i}^{\ast}}\cdot
\frac{2k_{iz}/k_{g}}{f_{z}+Z-i\tilde\delta}.
\end{equation}

Then the energy transmission coefficient $D_{i}^{\leftarrow}$ for the wave $i$
can be calculated as the fraction of the energy of the incident wave packet
that is transmitted into the solid.  It is explicitly shown that
\begin{equation}
     \label{Deq}
     D_{i}^{\leftarrow}(\chi,k_{\tau})=D_{i}^{\rightarrow}(\chi,k_{\tau}).
\end{equation}
This important relation ensures thermodynamic equilibrium between the solid and
helium at equal temperatures on both sides of the interface. Due to
Eq. (\ref{Deq}) we can from now on omit
the arrows in the sub- and superscripts of $D_{i}$ and $D$.

The reflection coefficients for $i\!=\!j$ are just $R_{ii}\!=\!|r_{ii}|^{2}$
and from Eq. (\ref{rii}) we obtain
\begin{equation}
     \label{Rii}
     R_{ii}=
     \left|\frac
     {(Z-i\tilde\delta)k_{g}f_{-2z}^{(i)}+f_{-3z}^{(i)}}
     {(Z-i\tilde\delta)k_{g}f_{2z}+f_{3z}}
     \right|^{2}.
\end{equation}
For $i\!\neq\!j$ we have to take into account that energy flows for all waves
are proportional to group velocities (\ref{U}), and then from Eqs.
(\ref{rij}) and (\ref{f-})
we derive
\begin{align}
     R_{ij}=R_{ji}&=
     4k_{g}^{2}
     \left|k_{iz}k_{jz}(k_{i}^{2}\!-\!k_{k}^{2})(k_{j}^{2}\!-\!k_{k}^{2})\right|\times\nonumber\\
     &\times\left|
     \frac{Z\!-\!i\tilde\delta+k_{kz}/k_{g}}
	{(Z\!-\!i\tilde\delta)k_{g}f_{2z}+f_{3z}}
     \right|^{2}.
     \label{Rij}
\end{align}
  The quantity $R_{ij}$ is the probability of quasiparticle $j$ being
created at the
interface when quasiparticle $i$ is incident, so the $R_{ij}$ can be
also called "conversion coefficients".

\begin{figure}[!bt]
\begin{center}
\includegraphics[viewport=105 110 500 400, width=0.45\textwidth]{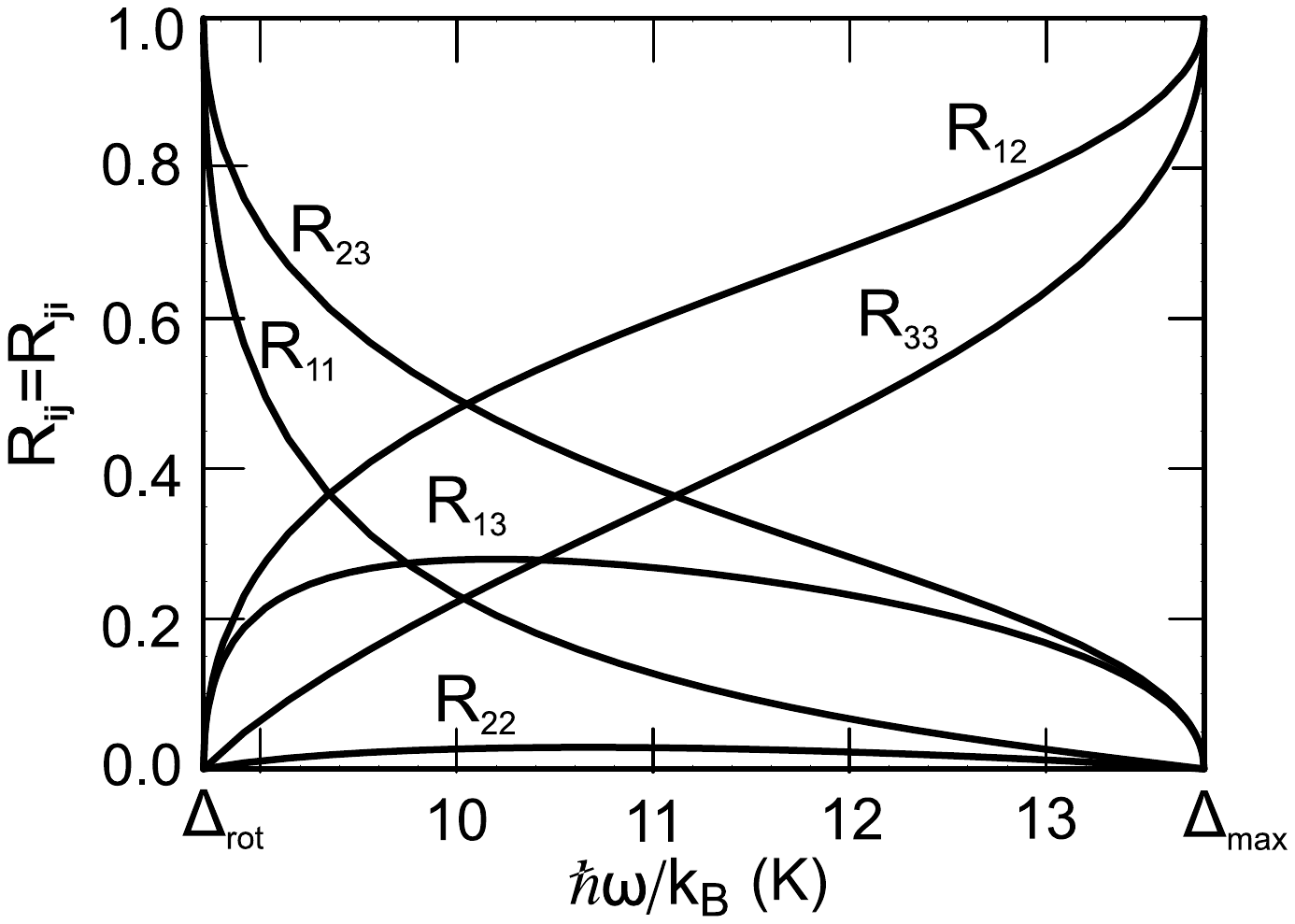}
\parbox{0.45\textwidth}
	{\caption{\label{Fig6} Functions $R_{ij}(\omega,\theta_{i}\!=\!0)$ for $i,j\!=\!1,2,3$.}}
\end{center}
\end{figure}

Their dependence on frequency at normal incidence is shown on Fig.\ref{Fig6}. We see,
in particular, that at the roton minimum $\chi\!\rightarrow\!\chi_{rot}$, where
the $R^-$ and $R^+$ roton branches merge, these quasiparticles are reflected
into each other with probability that tends to unity $R_{23}\!\rightarrow\!1$.
The same effect is present for phonons and $R^-$ rotons at the maxon maximum.

  The angular dependence of $R_{ij}$ is most easily analyzed in terms
of $k_{\tau}$
instead of the three angles of incidence. The values of $k_{\tau}$ equal to
$k_{sol}(\omega)$ or $k_{i}(\omega)$ correspond to different critical angles of
incidence. So, when $k_{\tau}\!\in\!(0,k_{sol})$, the quantities $Z$ and
$f_{\pm nz}^{(i)}$ (i.e.  $f_{nz}$ and  $f_{-nz}^{(i)}$ for $i\!=\!1,2,3$) are
all real, so all the waves are traveling waves and $D\!\neq\!0$. When
$k_{\tau}\!\in\!(k_{sol},k_{1})$, the wave in the solid is damped, $Z\!=\!i|Z|$
and $D\!=\!0$, but $f_{\pm nz}^{(i)}\!\in\!\mathbf{R}$ and all the waves in the
helium are still reflected into each other. When
$k_{\tau}\!\in\!(k_{1},k_{2})$, the phonon wave in helium is damped
$k_{1z}\!=\!i|k_{1z}|$ and no longer gives a traveling wave packet, and $f_{\pm
nz}^{(i)}$ also become complex. This corresponds to $R^{\pm}$ rotons incident
at angles greater than $\theta_{31,21}^{cr}$ and reflecting into themselves or
into each other. When $k_{\tau}\!\in\!(k_{2},k_{3})$ the quantities $f_{\pm
nz}^{(i)}$ are also complex but the structure is different; this case
corresponds to $R^{+}$ rotons incident at angles greater than
$\theta_{32}^{cr}$ and reflecting into $R^{+}$ rotons, again with probability
$1$.

In all the cases, energy conservation can be explicitly verified but it takes
different forms:
\begin{align}
     k_{\tau}\!<\!k_{sol}(\omega):&\;
         \sum\limits_{j=1}^{3}R_{ij}
	=1\!-\!D_{i}\quad\text{for}\;i\!=\!1,2,3;\nonumber\\
     k_{sol}(\omega)\!<\!k_{\tau}\!<\!k_{1}(\omega):&\;
         \sum\limits_{j=1}^{3}R_{ij}=1\quad\mbox{for}\;i\!=\!1,2,3;\nonumber\\
     k_{1}(\omega)\!<\!k_{\tau}\!<\!k_{2}(\omega):&\;
         R_{22}=R_{33}=1-R_{23};\nonumber\\
     k_{2}(\omega)\!<\!k_{\tau}\!<\!k_{3}(\omega):&\;
         R_{33}=1.
	     \label{En0}
\end{align}

\begin{figure}[t]
\begin{center}
\includegraphics*[viewport=105 0 485 250, width=0.45\textwidth]{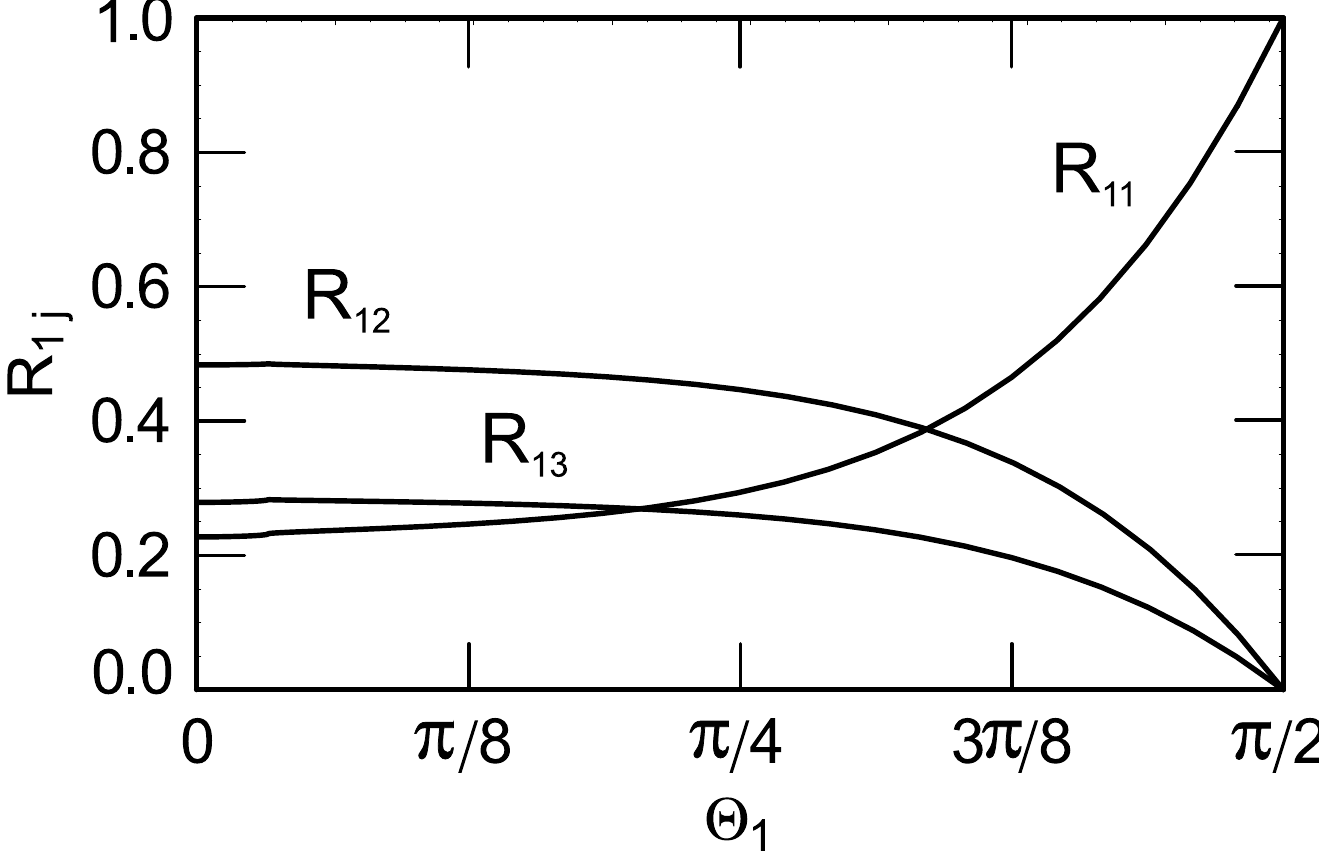}
\includegraphics*[viewport=105 0 485 250, width=0.45\textwidth]{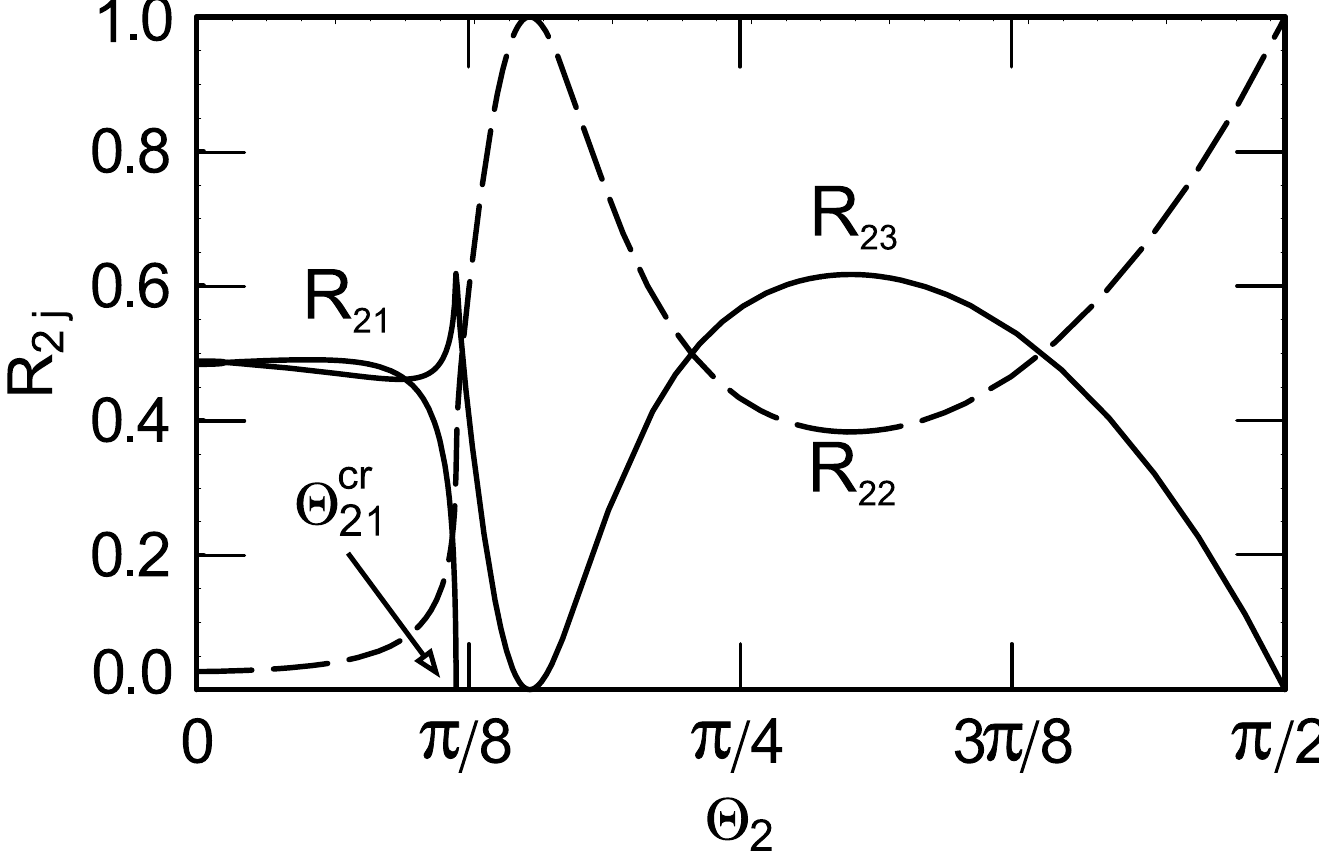}
\parbox{0.45\textwidth}
	{\caption{\label{Fig7a} \label{Fig7b} [(a) and (b)] Reflection coefficients $R_{1j}$ and $R_{2j}$ for $j\!=\!1,2,3$ as functions of incidence angle $\theta$ at $\chi\!=\!0.2871$ ($\hbar\omega/k_{B}\!=\!10K$).}}
\end{center}
\end{figure}

For the interface between helium and a solid, the limit $Z_{0}\!\ll\!1$ is a
good approximation, and in the Eqs. (\ref{rii}), (\ref{rij}) and (\ref{Rii}),
(\ref{Rij}) $Z$ can be neglected (in this limiting case $D\!\rightarrow\!0$ and
$\Theta_{i}^{cr}\!\rightarrow\!0$). However, the angles are not small anymore,
as was the case for $D_{i}$, and the angular dependence of the coefficients is
strong. This can be clearly seen in Fig.\ref{Fig7a}, where the graphs of $R_{1j}$ and
$R_{2j}$ are shown for $\hbar\omega/k_{B}\!=\!10K$. The coefficient $R_{21}$
becomes zero at the critical angle $\theta_{21}^{cr}$. The peak of $R_{22}$
and minimum of $R_{23}$ correspond to angles above critical, where $k_{1z}$
is imaginary and the damping depth of the phonon wave becomes roughly half of
the damping depth of the nonlocality kernel $h(r)$; then the
imaginary part of the
numerator of Eq. (\ref{Rij}) turns to zero, and as $Z$ is small, $R_{23}$ has a
deep minimum.

A more extensive analysis of the functions $R_{ij}(\chi,\theta_{i})$
for the case
$Z_{0}\!\ll\!1$ allows us to state the following: the main processes near the
roton minimum $\chi\!\rightarrow\!\chi_{rot}\!+\!0$, for all angles, are the
conversion of $R^-$ and $R^+$ rotons into each other and reflection of phonons
into themselves; near the maxon maximum phonons and
$R^-$ rotons are converted into each other and reflection of $R^+$
rotons into themselves.
For phonons and $R^-$ rotons, when the incidence angle becomes close
to $\pi/2$,
the probabilities of reflection into themselves $R_{11,22}$ tend to unity; for
$R^+$ rotons this happens at $\theta_{3}\!\rightarrow\!\theta_{32}^{cr}\!-\!0$,
and at greater angles $R_{33}\!=\!1$ exactly. The conversion coefficients
$R_{1j,j1}$ for $j\!=\!2,3$ are monotonically decreasing functions of
the  angles of incidence
; $R_{1j}$ becomes zero at $\theta_{j}\!\rightarrow\!\pi/2$ as
$\sqrt{\pi/2\!-\!\theta_{1}}$, $R_{j1}$ at
$\theta_{j}\!\rightarrow\!\theta_{j1}^{cr}$ as
$\sqrt{\theta_{j1}^{cr}\!-\!\theta_{j}}$. A little above $\theta_{j1}^{cr}$ the
coefficients $R_{22}$ and $R_{33}$ have high sharp peaks, and $R_{23}$ a
corresponding minimum, as described above.

Then for the case depicted in Fig.\ref{Fig4}, the most powerful beam will be
always beam
$ii$ (basically because of phonons with energies less than $\chi_{rot}$). We
have shown that $R^-$ rotons are hardly created by a solid heater
(\ref{Drot}), and the probability of $R^+$ rotons creation is also quite small
at frequencies near $\chi_{rot}$ (see Fig.\ref{Fig5}) if the incident
beam consists mainly of low energy phonons. It was shown in
\cite{phononpulses3a,phononpulses4} that in a phonon beam,  low
energy phonons (l-phonons) are converted into phonons with energy about $10K$
(h-phonons). The fraction of the energy  in the initial beam, which
is converted to the
h-phonons, can be up to $50\%$ \cite{phononpulses4}. The conversion coefficient
of these phonons to $R^-$ rotons is given by $R_{12}$ at
$\hbar\omega/k_{B}\!\approx\!10K$, it is much greater than at the roton minimum
and almost reaches $1/2$ at normal incidence, which is more than
$R_{11}$, see Fig.\ref{Fig7a}.

\begin{figure}[!b]
\begin{center}
\includegraphics[viewport=140 535 445 765, width=0.4\textwidth]{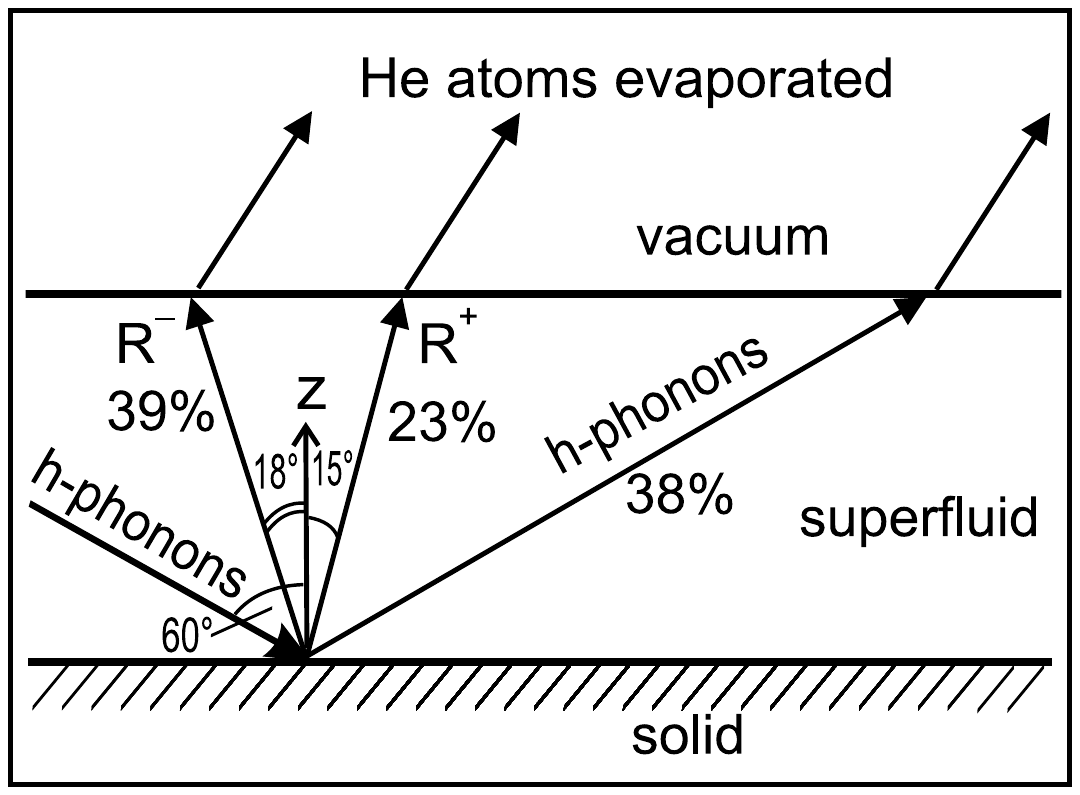}
\parbox{0.45\textwidth}
	{\caption{\label{Fig8} The predicted creation of $R^-$ rotons by  h-phonons incident on the interface with a solid. There should be backward reflection and quantum evaporation with backward refraction.}}
\end{center}
\end{figure}

We suggest the experimental setup depicted in Fig.\ref{Fig8}. The heater
injects a phonon beam, in which h-phonons are created. The h-phonons,
incident on the solid-helium interface,  are reflected into three beams of
phonons, $R^-$ rotons and $R^+$ rotons of comparable intensities (the $R^-$
rotons are reflected backwards). These beams propagate towards the free surface
of helium, and quantum evaporate atoms from it (the $R^-$ rotons
evaporate atoms
backward \cite{exp2}), which are then detected. Thus the energy is transported
from the heater to the  interface by phonons, and then to the detector by $R^-$
rotons along
a Z-shaped trajectory, with retro-reflection at the point of creation of $R^-$
rotons and retro-refraction on the surface. The angles and fractions of the
initial beam's energy, which is transferred to different reflected
beams, are shown for
the h-phonon part of the incident beam. The l-phonons for the most part are
directly reflected and are not shown.

If the source of quasiparticles has more rotons in the
incident beam, as the one used in \cite{exp2}, then beams $32$ and
$23$ may become also be detectable.

The main contribution to the energy flow through the interface at low
temperatures can be expected to be made by  phonons below the roton gap, i.e.
with $\chi\!\in\!(0,\chi_{rot})$. The problem of transmission through
interfaces by phonons with anomalous dispersion was solved in Refs.
\cite{PhNT} and \cite{JLTP2006}. In the current work, the dispersion relation
(\ref{OmegaRot}) that is used is more general than the one used in the
previous works, it is non-monotonic and normal below the roton gap.

When $\chi\!<\!\chi_{rot}$, the roots $k_{2z,3z}$ are defined so that
$k_{1z}\!>\!0$ and $k_{3z}\!=\!-k_{2z}^{\ast}\!\in\!\mathbf{C}_{+}$, as shown
in section 2.3. With these $k_{iz}$ the out- and in-solutions are constructed
(\ref{out-solutions}), (\ref{in-solutions}). So the amplitude coefficients are
still defined by Eqs. (\ref{r->}) and (\ref{ti}), but the quantities $f_{\pm
nz}^{(i)}$ are now complex. The only valid reflection coefficient $R_{11}$ is
defined by (\ref{Rii}) with the complex $k_{iz}$ as introduced above. The
transmission coefficient is $D_{ph}\!=\!1\!-\!R_{11}$. It can be shown that in
the limit of small frequencies $\chi\!\rightarrow\!0$, when the dispersion is
almost linear, the expression for $D_{ph}$ approaches the standard one for
linear dispersion. At $\chi\!\rightarrow\chi_{rot}\!-\!0$ it decreases rapidly
to less than half because of the increasing influence of the roton waves. The
curve $D(\chi)$ is continuous at $\chi_{rot}$ but has a kink.

\section{Energy flows through the interface}

When a phonon in the solid, of frequency $\omega$ and wave vector
$\mathbf{k}_{sol}$,
is incident on the interface at angle $\theta_{sol}$, the average energy
transferred into helium is $\hbar\omega D(\omega,k_{\tau})$, where $D$ is given
by (\ref{D->}) and $k_{\tau}\!=\!k_{sol}\cos\theta_{sol}$. Let the
phonons in the solid be
in thermodynamic equilibrium at temperature $T$. Then the normal component of
the density of energy flow through the interface is
   (see for example
\cite{Khalatnikov})
\begin{equation}
     \label{Q_Khalat}
     Q(T)=\int\frac{d^{3}k_{sol}}{(2\pi)^{3}}
     \hbar\omega \;n_{T}(\omega) s_{sol}\cos{\theta_{sol}}\,D,
\end{equation}
where $n_{T}$  is the Bose-Einstein distribution function and the
integration domain is
the half-space $k_{sol\,z}\!>\!0$. The parts of this energy flow, that are
transferred into helium by either phonons, or $R^-$, or $R^+$ rotons of helium
that are created at the interface by the incident phonons, are obtained in the
same way. But instead of the full
coefficient $D$, we now use the partial transmission coefficients
$D_{i}$. These are the
corresponding creation probabilities of the quasiparticles.
After changing the integration variables to the arguments of
$D_{i}(\omega,k_{\tau})$, the partial energy flows can be expressed in the form
\begin{equation}
     \label{flows}
     Q_{i}^{\rightarrow}(T)=\int
     \frac{d\omega}{8\pi^2}
     \hbar\omega \;n_{T}(\omega)\!\!
     \int\limits_{0}^{k_{i}^{2}(\omega)}\! dk_{\tau}^{2}D_{i}(\omega,k_{\tau}).
\end{equation}
Here the upper limit by $k_{\tau}^{2}$ corresponds to the maximum transmission
angle of quasiparticles of type $i$, equal to $\theta_{i}^{cr}$ from
(\ref{CritAnglesSolid}). The quantities $Q_{i}^{\rightarrow}$ for
$i\!=\!1,2,3$,
are the individual contributions of phonons, $R^-$, and $R^+$
rotons, to the energy flux from the solid into helium.

Their contributions to the energy flux, in the opposite direction
$Q_{i}^{\leftarrow}(T)$, are the normal components of the energy
fluxes from helium
into the solid. These are realized by helium quasiparticles of type
$i$ incident
on the interface. The average energy transferred into the solid per incident
quasiparticle of type $i$ is, due to (\ref{Deq}), $\hbar\omega
D_{i}(\omega,k_{\tau})$. Then the energy flux is derived in the same way as Eq.
(\ref{Q_Khalat}), with the difference that instead of $s_{sol}$ in the integral
we have $|u_{i}|$, because the number of quasiparticles incident on the
interface per unit of time is proportional to their group velocity:
\begin{equation}
     \label{Q_Back}
     Q_{i}^{\leftarrow}(T)=\int\frac{d^{3}k_{i}}{(2\pi)^{3}}
     \hbar\omega \;n_{T}(\omega) |u_{i}|\cos{\Theta_{i}}\,D_{i}.
\end{equation}
When changing the integration variables to $(\omega,k_{\tau})$,
we use the Jacobian determinant and we obtain explicitly
that $Q_{i}^{\leftarrow}(T)\!=\!Q_{i}^{\rightarrow}(T)$.

\begin{figure}[ht]
\begin{center}
\includegraphics[viewport=105 295 495 545, width=0.45\textwidth]{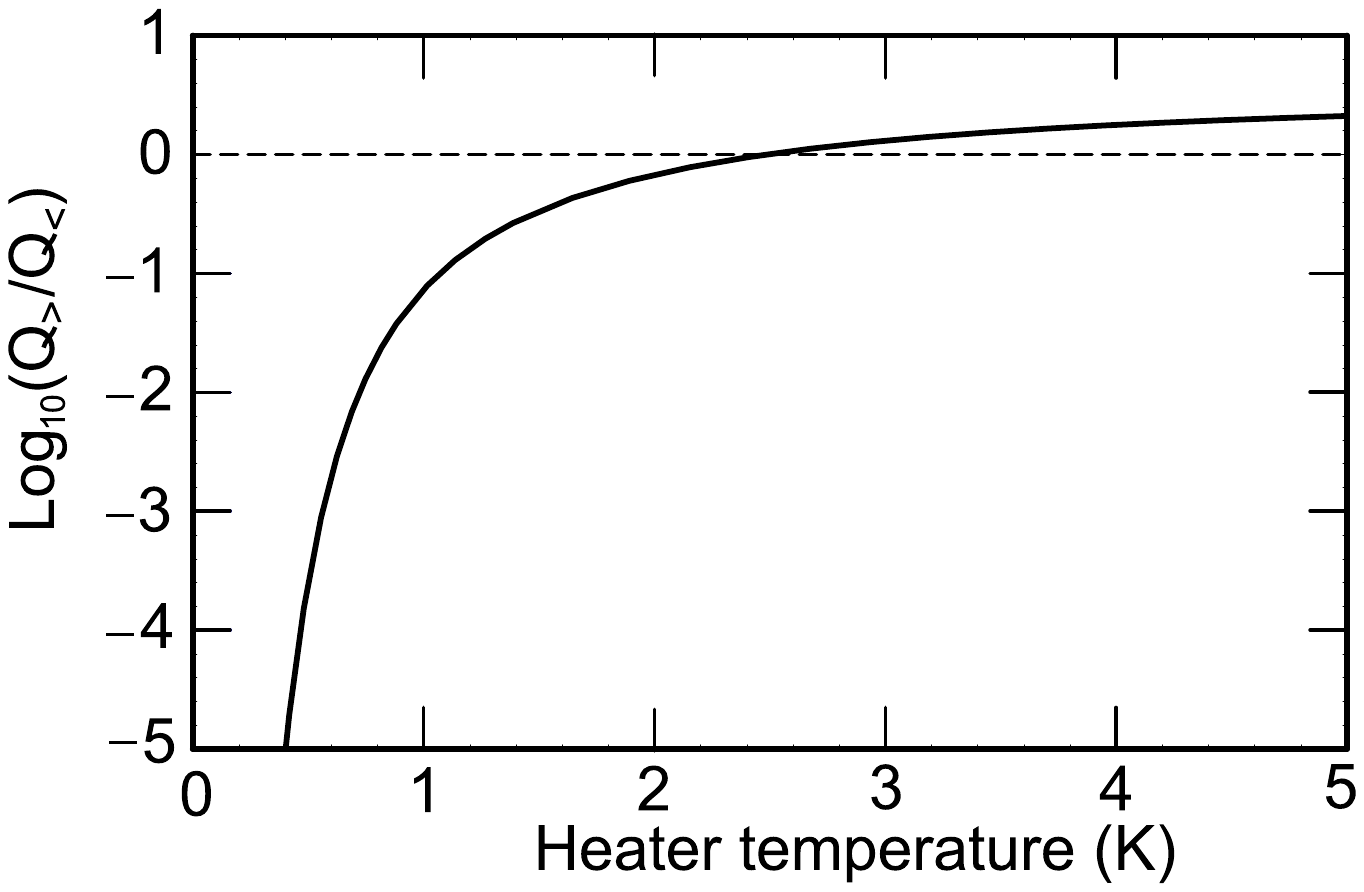}
\includegraphics[viewport=90 0 515 280, width=0.45\textwidth]{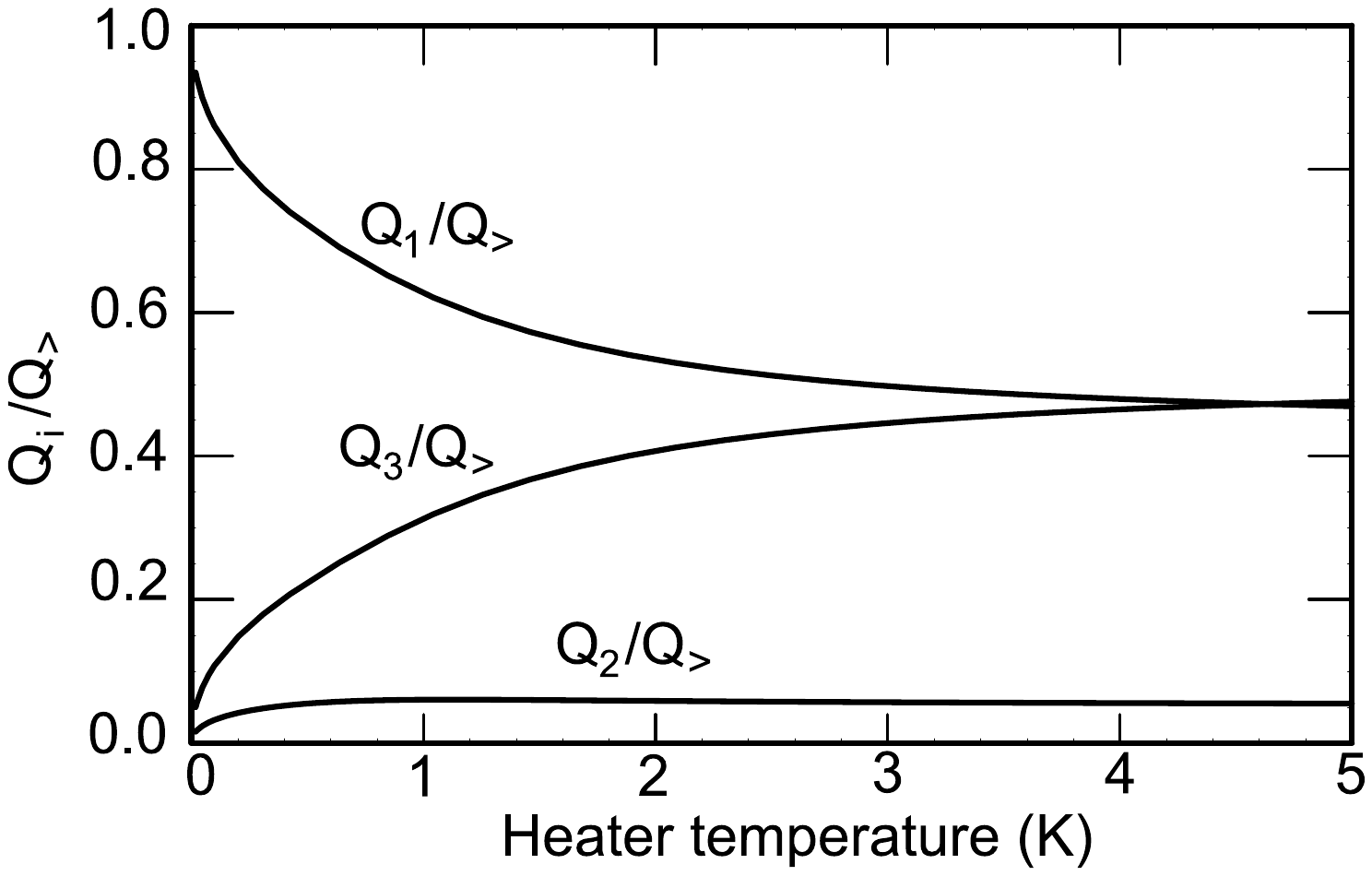}
\parbox{0.45\textwidth}
	{\caption{\label{Fig9} (a) The ratio of the contributions to the energy flow through the interface by quasiparticles above roton minimum, $Q_{>}$, to the phonons below roton minimum, $Q_{<}$, as a function of temperature (on a logarithmic scale). (b) The contributions of phonons, $R^-$ rotons and $R^+$ rotons to the energy flow created by quasiparticles above the roton minimum, as
functions of $T$}}
\end{center}
\end{figure}

Fig.\ref{Fig9}a shows the ratio between the contributions to the energy flux through
the interface of all quasiparticles above the roton gap (i.e. with
$\hbar\omega/k_{B}\!>\!\Delta$) $Q_{>}\!=\!Q_{1}\!+\!Q_{2}\!+\!Q_{3}$, and
the contribution of phonons below the roton gap $Q_{<}$, which is
obtained using (\ref{flows}) with $D_{ph}$. We see that at
temperatures $T\!<\!1$ K the phonons are dominant. However, at
$T\!\approx\!2.5K$ the two contributions are equal, and at higher temperatures
the phonons and rotons above the roton gap play the main role in heat exchange
with the solid, see Fig.\ref{Fig9}b. The contribution of the $R^+$ rotons  to
$Q_{>}$ increases with
temperature and at $T\!\approx\!3$ K their contribution surpasses
that of the phonons, see Fig.\ref{Fig9}b.
The  contribution of the $R^-$
rotons,  is approximately constant and is no greater than $6\%$.
This is due to the low creation probability $D_{2}$ for all
frequencies (\ref{Drot}).
At $T\!=\!3$ K the contribution of the $R^-$ rotons, to the full
energy flow, is $3\%$.

When there is energy flow through the interface, it induces the Kapitza
temperature jump at the interface (see for example \cite{Khalatnikov}). The
contributions of quasiparticles of each type to this jump are obtained by
differentiating Eq. (\ref{flows}) with respect to $T$.

\section{Conclusion}

In this work we have solved the problem of the interaction of $HeI\!I$
quasiparticles, i.e. phonons, $R^-$ rotons, and $R^+$ rotons, with
the interface between
helium and a solid. These excitations have
the non-monotonic dispersion curve shown in (Fig.\ref{Fig1}).
  The consistent solution of the problem has been introduced,
which allows us to rigorously describe the simultaneous creation of
the three types
of $HeI\!I$ quasiparticles by any one of them, or by a phonon in the
solid which is incident
on the interface.

When a phonon in the solid is incident on the interface, it is
reflected with some
probability and a phonon, $R^-$ roton, or $R^+$ roton are created
with the corresponding probabilities, in the helium. It is shown that
the created $R^-$ roton,
due to its negative group velocity, is refracted backward (Fig.\ref{Fig3}). When some
quasiparticle of helium is incident, all the quasiparticles, with the same
frequency and transverse wave vectors, are created. The set of six critical
angles as functions of frequency are introduced
(\ref{CritAnglesSolid}), (\ref{CritAnglesHe}). These separate the intervals of
angles of incidence for the different quasiparticles,
from which other quasiparticles can be created. It is shown when a
phonon or $R^+$ roton is
incident, the $R^-$ roton is retro-reflected (i.e. reflected backwards), and
likewise when an $R^-$ roton is incident, the phonon and $R^+$ rotons are
retro-reflected (Fig.\ref{Fig4}). This effect is the Andreev reflection of phonons and
rotons.

The probabilities of creation of all quasiparticles at the interface,
when any quasiparticle is incident,
  are derived as functions of frequency and incidence
angles (\ref{D->i}), (\ref{Rii}), (\ref{Rij}) and Figs.\ref{Fig5},\ref{Fig6},\ref{Fig7a}. It is shown that
the creation probability of an $R^-$ roton by a phonon in the solid,
and vice versa,
is very small for all angles and frequencies (\ref{D2appr}), (\ref{Drot}). This
means that $R^-$ rotons are as badly created by a solid heater as they
are poorly detected by a solid bolometer. This explains the
failure to detect $R^-$ rotons in direct experiments until 1999
\cite{exp2}. New predictions are made for experiments with beams
of phonons and rotons interacting with the solid interface, and in
particular, creating  $R^-$ rotons at
the interface by a beam of high-energy phonons (h-phonons).

The full energy flow through the interface, is also
calculated as a function of temperature of  the solid, as well as the
individual contributions
of the phonons, $R^+$, and $R^-$ rotons (\ref{flows}) to it, see Fig.\ref{Fig9}.
The contribution of the $R^-$ rotons is shown to be very small.

\acknowledgements{
We are grateful to  Adrian Wyatt for many useful discussions and to EPSRC of the UK (grant EP/F 019157/1) for support of this work.}

\end{document}